\documentclass[sigconf,nonacm]{acmart}

\usepackage{xcolor}
\usepackage{tcolorbox}
\usepackage{enumitem}
\usepackage{booktabs}
\usepackage[table]{xcolor}
\usepackage{graphicx}
\usepackage{multirow}
\usepackage{bbold}
\usepackage{algorithm}
\usepackage[noabbrev]{cleveref}
\usepackage{listings}
\usepackage{fontawesome5}

\definecolor{codebackground}{RGB}{248,248,248}
\definecolor{codecomment}{RGB}{35,110,55}
\definecolor{codekeyword}{RGB}{35,70,150}

\lstdefinestyle{pytorch}{
      language=Python,
      basicstyle=\ttfamily\footnotesize,
      keywordstyle=\color{codekeyword},
      commentstyle=\color{codecomment},
      frame=none,
      columns=fullflexible,
      keepspaces=true,
      showstringspaces=false,
      breaklines=true,
      breakatwhitespace=true,
      numbers=left,
      numberstyle=\tiny\color{black!45},
      numbersep=5pt,
      xleftmargin=3.5em,
      framexleftmargin=1em,
}

\lstdefinestyle{pytorch-muted}{
    style=pytorch,
    basicstyle=\ttfamily\footnotesize\color{black!50},
    keywordstyle=\color{codekeyword!50},
    commentstyle=\color{codecomment!50},
    stringstyle=\color{black!50},
    numberstyle=\tiny\color{black!100},
    aboveskip=0pt,
    belowskip=0pt,
}

\lstdefinestyle{pytorch-focus}{
    style=pytorch,
    basicstyle=\ttfamily\footnotesize\color{black},
    keywordstyle=\color{codekeyword},
    commentstyle=\color{codecomment},
    stringstyle=\color{black},
    numberstyle=\tiny\color{black!100},
    aboveskip=0pt,
    belowskip=0pt,
}

\newcommand{\R}{\mathbb{R}}

\newcommand{\trans}{\mathsf{T}}

\graphicspath{{FIGURES/}}
\newcommand{\sid}[1]{\texttt{#1}}
\DeclareMathOperator*{\argmin}{arg\,min}

\AtBeginDocument{%
  }

\usepackage{array}
\usepackage{enumitem}

\newlist{nearestitems}{enumerate}{1}
\setlist[nearestitems]{
    label=(\arabic*),
    leftmargin=1.7em,
    itemsep=0pt,
    topsep=0pt,
    parsep=0pt,
    partopsep=0pt
}

\begin{document}

\title{Preserving Item Semantics for Free: Rethinking Token Initialization in LLM-Based Generative Recommendation}

\author{Donald Loveland}
\affiliation{%
  \institution{Snap Inc.}
  \city{Bellevue, WA}
  \country{USA}
}
\email{dloveland@snapchat.com}

\author{Liam Collins}
\affiliation{%
  \institution{Snap Inc.}
  \city{Bellevue, WA}
  \country{USA}
}
\email{lcollins2@snapchat.com}

\author{Bhuvesh Kumar}
\affiliation{%
  \institution{Snap Inc.}
  \city{Bellevue, WA}
  \country{USA}
}
\email{bkumar4@snapchat.com}

\author{Danai Koutra}
\affiliation{%
  \institution{University of Michigan}
  \city{Ann Arbor, MI}
  \country{USA}
}
\email{dkoutra@umich.edu}

\author{Neil Shah}
\affiliation{%
  \institution{Snap Inc.}
  \city{Bellevue, WA}
  \country{USA}
}
\email{nshah@snapchat.com}
\renewcommand{\shortauthors}{Loveland et al.}

\begin{abstract}
    Recent advances in generative recommendation (GR) leverage large language models (LLMs) as recommender backbones, enabling LLMs to directly generate recommendations conditioned on item-interaction histories. In these systems, items are often represented through semantic IDs (SIDs) added to the LLM vocabulary as special tokens. Ideally, SIDs imbue item token representations with semantic priors, thereby improving model generalization. However, standard vocabulary expansion typically initializes these tokens as random Gaussian vectors, discarding the SIDs' underlying continuous geometry and forcing the LLM to relearn token relationships from interaction data. To demonstrate the consequences of this design, we first show that training from this initialization tends to organize SID embeddings around item popularity rather than semantics. We further show that, despite partially reducing the reliance on popularity and improving cold item performance, the computationally expensive process of continual pretraining (CPT) fails to reliably recover the original semantic geometry. To address these findings, we propose a simple, parameter-free intervention that initializes SID token embeddings directly from their corresponding centroids in the semantic embedding space. Requiring only a few lines of code and no additional training or inference overhead, this drop-in approach improves pure-SFT \textbf{Recall@5 by up to 16\%}, reaches peak performance with \textbf{up to 40\%} fewer SFT steps, and improves \textbf{cold-item Recall@5 by up to 60\%}. Moreover, on datasets that benefit from additional CPT, centroid initialization reaches comparable performance while \textbf{requiring half as many CPT epochs}. Together, our findings show that preserving SID geometry, beyond shared-prefix structure, provides a simple and effective semantic prior for LLM-based GR. 
\end{abstract}



\maketitle

\section{Introduction}

Recommender systems play a central role in connecting users with relevant items across large and complex catalogs \cite{gomez2016netflix, liu2019recsyspersonal, fan2019graphrec, schafer1999recec, jin2024amazon}. With the rise of transformer architectures, recommendation has increasingly shifted toward generative recommendation (GR), a formulation in which transformer-based models autoregressively generate items conditioned on item-interaction histories \cite{rajput2023recommender, wang2024eager, zhai2024actions, deng2025onerecunifyingretrieverank}. Building on this paradigm, advances in pretrained LLMs have motivated their adoption as generative backbones, giving rise to LLM-based GR \cite{hou2024large, MiniOneRec, liu2025onerecthinkintextreasoninggenerative, li2024large}. This extension promises to combine collaborative signals from item interactions with semantic knowledge acquired during LLM pretraining \cite{zheng2024adapting, yang2025qwen3technicalreport, dong2025ctp}. Realizing this promise, however, is nontrivial, and recent research and large-scale systems have adopted increasingly complex, multi-stage training pipelines to align item representations with the LLM while preserving the collaborative structure needed for accurate recommendation \cite{liu2025onerecthinkintextreasoninggenerative,chen2026groundedtokeninitializationnew, Lin2026tcrec, wang2024learnable}.

Central to LLM-based GR is the interface between catalog items and the LLM token space \cite{geng2022recommendation, hou2026surveygenerativerecommendationdata}. Importantly, this interface must allow the LLM to encode item histories and generate valid items without incurring catalog-scale memory and inference costs \cite{qu2025tokenrec}. Semantic IDs (SIDs) have emerged as the standard solution, representing each item as a short sequence of discrete codes obtained by quantizing a content-derived item embedding \cite{hou2023learning, singh2024better, rajput2023recommender, ju2025generative}. The standard LLM-based GR pipeline exposes these identifiers to the LLM by adding each SID code to its vocabulary as a special token, commonly referred to as vocabulary expansion \cite{liu2025onerecthinkintextreasoninggenerative, zheng2024adapting, he2026reasoning}. Through the composition of a small number of codes across hierarchical levels, SIDs can encode extremely large sets of items with only modest vocabulary expansion, e.g., three levels with 256 codes each can uniquely express up to \(16.8\) million items while adding only \(768\) tokens. 
Therefore, SIDs effectively decouple the LLM vocabulary size from the total number of items, satisfying the memory and computational constraints required for scalable GR.

\begin{figure}[t]
\centering
\includegraphics[width=\columnwidth]{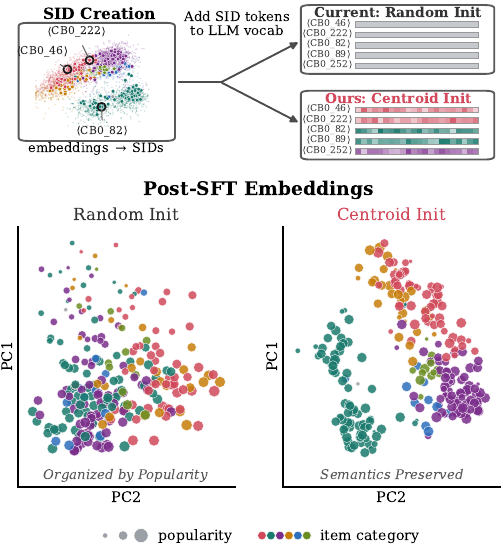}
\caption{Learned embedding geometry for Amazon Beauty dataset after SFT. \emph{Top:} Standard LLM vocabulary expansion discards the semantic geometry, whereas our proposed \textit{centroid initialization} preserves it. \emph{Bottom:} After SFT, randomly initialized embeddings strongly encode popularity, while centroid-initialized embeddings retain semantics.}
\vspace{-0.3cm}
\label{fig:teaser}
\Description{Standard vocabulary expansion discards SID centroid geometry, while centroid initialization transfers it to the token embeddings. After SFT, randomly initialized embeddings align with popularity, whereas centroid initialized embeddings retain semantics. Colors indicate categories and sizes indicate code popularity.}
\end{figure}

Beyond vocabulary expansion, standard LLM-based GR relies on multi-stage training to integrate the newly added SID tokens. This recipe is built on two foundational phases: (a) continual pretraining~(CPT), also sometimes referred to as itemic alignment (IA), 
to ground the SID embeddings in the natural language space, and (b)~supervised fine-tuning (SFT) to teach the recommendation task~\cite{liu2025onerecthinkintextreasoninggenerative, he2026plum, zheng2024adapting, he2026implicitreasoninglargelanguage, he2026reasoning}.\footnote{While some pipelines employ subsequent post-training for task-specific goals (e.g., reinforcement learning to elicit thinking \cite{Yu_2026, liu2025onerecthinkintextreasoninggenerative}), we focus on CPT and SFT as they constitute the foundational phases that establish the underlying item representations.} 
However, an overlooked aspect of this pipeline is the initialization of the SID token embeddings, which we posit dictates how these representations evolve during training and impacts their downstream performance.
Ideally, SIDs should imbue the token representations with rich semantic priors from the moment they are introduced \cite{penha2025semantic}. Yet, standard vocabulary expansion does not support this goal \cite{Dobler_2023}; instead, it initializes new SID tokens as random Gaussian vectors whose mean and covariance match the pre-existing word tokens. 
While this strategy keeps the new tokens in-distribution of the vocabulary, it actively discards the original semantic geometry established at SID construction, forcing the model to relearn these relationships from interaction data during CPT and SFT \cite{chen2026groundedtokeninitializationnew, he2026plum, liu2025onerecthinkintextreasoninggenerative}. This raises our first research question: \textit{In the absence of their original geometry, what do the SID embeddings learn, and are the underlying semantics recovered?}

To answer this question, we start with a series of diagnostic analyses to examine what signals SID embeddings capture. Across both SFT and CPT+SFT, we find that the semantic geometry, discarded during vocabulary expansion, is \textit{not} recovered. Even more concerning, we demonstrate that the SID embeddings tend to become organized primarily by popularity (as shown in \Cref{fig:teaser}). We mathematically explain this phenomenon through a gradient analysis, and also demonstrate that this learned structure reinforces downstream popularity bias. Furthermore, while we show that CPT can partially dilute this reliance on popularity and improve tail-item recall, CPT still fails to reliably recover the original semantic geometry while incurring a large training overhead \cite{he2026plum, liu2025onerecthinkintextreasoninggenerative, chen2026groundedtokeninitializationnew}. These findings together reveal that standard pipelines leave the continuous semantic prior of the SIDs largely underutilized. More broadly, by defaulting to simpler signals like popularity, these systems regress to inheriting the well-known biases of traditional recommenders \cite{loveland2025role, mansoury2020recsys, lin2025resn}, undermining the goal of better semantic utilization provided by LLMs \cite{singh2024better, liu2023text, xi2024towards}.

To overcome this limitation, we pose our second research question: \textit{How can we enable an LLM-based GR model to directly exploit the continuous semantic structure of its SIDs?} Our answer focuses on simplicity: rather than relying on costly CPT to recover semantics under standard random initialization, we inject the SIDs' underlying semantic geometry directly into the LLM. 
Specifically, we propose a straightforward strategy of initializing each SID token's embedding directly from its corresponding centroid in the original semantic space.
Acting as a direct bridge between the item semantics and the LLM, this intervention incurs zero additional training or inference overhead while requiring only a few lines of code. Across five datasets, centroid initialization improves pure-SFT Recall@5 by 6.5\% while reducing the number of SFT steps by 27\% on average. Additionally, these gains are largest on cold items, where Recall@5 improves by 37\% on average without CPT. Furthermore, on datasets where random initialization benefits from deeper CPT, centroid initialization reaches comparable performance with only 1-4 CPT epochs (rather than 6-10). More broadly, we argue that preserving the geometry of the semantic embeddings moves the field closer to the true promise of LLM-based GR, enabling recommenders to more directly leverage knowledge of item content, rather than overindexing on signals like popularity. Our contributions are below:

\begin{itemize}[leftmargin=1.4em,itemsep=2pt]
      
\item \textbf{Diagnosing SID Embedding Properties.}
  We uncover a previously overlooked training dynamic in LLM-based GR models where the SID embeddings discard their original semantic geometry, organizing instead around item popularity. We then explain how random initialization promotes the encoding of popularity and show its influence on popularity bias.
\item \textbf{Geometry-Preserving Vocabulary Expansion.}
We introduce centroid initialization, a parameter-free intervention that transfers SID centroids directly into the corresponding SID token embeddings. As a drop-in replacement for standard initialization, it preserves SID geometry through a one-time embedding assignment, requiring no auxiliary training or architectural changes.

    \item \textbf{Empirical Benefits and Regimes of Effectiveness.}
Across datasets, centroid initialization improves pure-SFT Recall@5 by 6.5\%, reaches peak performance with 27\% fewer SFT steps, and improves cold-item Recall@5 by 37\% on average. On the datasets that benefit from deeper CPT, it also reaches comparable performance with 1-4 CPT epochs rather than 6-10, while retaining a 14\% cold-item advantage after one epoch.

\end{itemize}

\section{Preliminaries and Notation}\label{sec:setup}

In this section, we formalize the LLM-based GR framework used throughout the paper. At a high level, the core objective of LLM-based GR is to predict a user's next item based on their historical interactions. Formally, given a sequence of item interactions $H = (i_1, \dots, i_{|H|})$, the model aims to generate the subsequent item $i_{|H|+1}$. To ensure the model generates valid items without causing an explosion in vocabulary size, SIDs are utilized both to encode the interaction history and to define the generation target. The remainder of this section details SID construction, standard vocabulary expansion, training objectives, and the inference process. While we discuss some related work in this section, further details are provided in \Cref{sec:related}. 

\vspace{0.1cm}
\noindent \textbf{Semantic IDs (SIDs).}
To generate SIDs, we first assume each item~$i$ in the catalog is represented by a continuous semantic embedding $\mathbf{z}_i\in\mathbb{R}^{d}$, typically extracted from a pretrained encoder. To map these vectors into discrete token sequences, we apply residual $k$-means quantization with $L$ semantic levels and $K$ codes per level. At each level $\ell\in\{0,\dots,L-1\}$, let
$\mathcal{C}_{\ell}=\{\mathbf{c}_{\ell,k}\}_{k=0}^{K-1}$ denote the level-$\ell$ codebook, where $\mathbf{c}_{\ell,k}\in\mathbb{R}^{d}$ is the centroid indexed by code $k$. 
Then, for an embedding $\mathbf{z}_i$, the residual is initialized as $\mathbf{r}_0 = \mathbf{z}_i$. Once initialized, the following recursive process is run at each level $\ell$: the quantizer normalizes the current residual, assigns it to the nearest centroid, and computes the new residual for the next level. This is expressed as,
\begin{equation}
    \mathbf{u}_\ell = \frac{\mathbf{r}_\ell}{\lVert \mathbf{r}_\ell\rVert_2}, \qquad k_\ell(i) = \underset{k\in\{0,\dots,K-1\}}{\argmin}\,\lVert \mathbf{u}_\ell-\mathbf{c}_{\ell,k}\rVert_2^2, \qquad \mathbf{r}_{\ell+1} = \mathbf{u}_\ell-\mathbf{c}_{\ell,k_\ell(i)},
    \label{eq:rq}
\end{equation}
producing a sequence of codes $k_0(i), \dots, k_{L-1}(i)$. The level-0 code $k_0(i)$ captures the item's coarse semantics, while subsequent codes capture increasingly fine-grained details. However, due to the discrete nature of the quantization space, multiple items with similar semantics can map to the same sequence of codes. To disambiguate these collisions, we append an additional deduplication code $k_L(i)$. Unlike the preceding codes, this final identifier is not associated with a semantic centroid. 
The resulting SID for item $i$ is the ordered tuple of these codes:
$$
\mathrm{SID}(i) = \big(k_0(i), \dots, k_{L-1}(i), k_L(i)\big).
$$
To process these sequences, each code index $k$ at level $\ell$ is mapped to a discrete token in the LLM's expanded vocabulary.

\vspace{0.1cm}
\noindent \textbf{Vocabulary Expansion and Initialization.}
To allow the LLM to natively process and generate SIDs, the discrete codes representing the items must be added to the model's vocabulary. Let $\mathcal{V}_{\mathrm{base}}$ denote the original vocabulary, $\mathcal{V}_{\mathrm{new}}$ the newly added tokens, and $\mathcal{V}_{\mathrm{SID}}\subseteq\mathcal{V}_{\mathrm{new}}$ the subset corresponding to SID codes. The expanded vocabulary is $\mathcal{V}=\mathcal{V}_{\mathrm{base}}\cup\mathcal{V}_{\mathrm{new}}$. Let $\mathbf{E}\in\mathbb{R}^{|\mathcal{V}|\times d_{\text{model}}}$ denote the LLM's token embedding matrix, which maps discrete tokens into continuous $d_{\text{model}}$-dimensional representations (where the LLM hidden dimension $d_{\text{model}}$ may differ from the original semantic embedding dimension $d$). Expanding the vocabulary entails appending a new row $\mathbf{e}_t\in\mathbb{R}^{d_{\text{model}}}$ to $\mathbf{E}$ for each token $t\in\mathcal{V}_{\mathrm{new}}$. 

A common vocabulary expansion strategy in LLMs is to initialize newly added token embeddings to match the distribution of the pretrained vocabulary embeddings \cite{yamaguchi2026, mundra2024empirical}. This strategy is designed to minimize disruption with respect to the model's pre-training. In LLM-based GR, a similar practice is typically followed \cite{liu2025onerecthinkintextreasoninggenerative, he2026plum, he2026implicitreasoninglargelanguage}. Specifically, let $\boldsymbol{\mu}_{\mathrm{base}}$ and $\boldsymbol{\Sigma}_{\mathrm{base}}$ denote the empirical mean and covariance of the embeddings for $\mathcal{V}_{\mathrm{base}}$. Newly added SID embeddings are then initialized by sampling from a scaled normal distribution:
\begin{equation}
\mathbf{e}_t \sim \mathcal{N}(\boldsymbol{\mu}_{\mathrm{base}},\varepsilon\boldsymbol{\Sigma}_{\mathrm{base}}),
\label{eq:randinit}
\end{equation}
where $\varepsilon$ is a small variance scaling factor. 
While this strategy, which we refer to throughout the paper as \textit{random initialization}, is reasonable for general NLP tasks, we argue that it introduces a potential bottleneck for GR. Given the random Gaussian vectors are agnostic to the items' semantic meaning, the  SIDs are initially treated as arbitrary tokens. As a result, the embeddings must reconstruct the semantic structure entirely from scratch during training, relying solely on patterns like shared prefixes or interaction co-occurrence. 

\vspace{0.1cm}
\noindent \textbf{Training Stages.}
For a given an item-interaction history $H$ and a target item $i$ represented by its SID sequence $y=(y_0,\dots,y_L)$, where $y_j=k_j(i)$, the recommendation model is trained to minimize the standard autoregressive next-token prediction loss:
\(
\mathcal{L}(\theta) = -\sum_{j=0}^{L} \log p_\theta\!\big(y_j \mid y_{<j}, H\big).
\label{eq:loss}
\)
As the newly added SID tokens lack a meaningful initial geometry, standard LLM-based GR pipelines optimize this objective through a two-stage training process. At a high-level, the first stage is designed to encourage the LLM to relearn the semantic structure, whereas the second stage trains the LLM to perform next item prediction. Details on these stages are below, with prompts detailed in \Cref{sec:prompts}:
\begin{enumerate}[leftmargin=*]
    \item \textbf{Continual Pre-Training (CPT):} As the standard initialization discards the items' original structure, CPT aims to enable the relearning of these relationships.
    Typically, during CPT, the LLM backbone is frozen, and only the newly added token embeddings ($\theta_{\mathrm{CPT}}=\{\mathbf{e}_t:\,t\in\mathcal{V}_{\mathrm{new}}\}$) are updated. Learning occurs over SID-text pairs with the goal of predicting the text of an item given its SID, effectively grounding the SIDs in natural language. 
    \item \textbf{Supervised Fine-Tuning (SFT):} Once the token embeddings are sufficiently warmed up, SFT aims to enable the LLM to perform recommendation. In this phase, all model parameters $\theta$ are unfrozen and updated. Training utilizes item interactions, tasking the model to autoregressively predict the target item's SID conditioned on a historical context of previously interacted SIDs. 
\end{enumerate}
Throughout the paper, we refer to training with the second stage alone as \emph{pure SFT}, and the application of both stages as \emph{CPT+SFT}.

\vspace{0.1cm}
\noindent \textbf{Generation.}
The model generates recommendations by sampling from $p_\theta(\cdot \mid H)$. However, unconstrained generation risks generating sequences of SID tokens that do not map to an existing item. To ensure that the generated sequences correspond to valid items, we enforce constrained generation. Specifically, at each generation step $j$, the vocabulary is masked such that the probability of tokens that are not valid children of the previously generated prefix $y_{<j}$ is set to zero. We apply this constrained beam search to generate multiple valid SIDs, then rank the SIDs by their log-probability to output the top-$k$ items.

\section{What do SID Embeddings Encode in Standard LLM-based GR Pipelines?}
\label{sec:motivation}

In this section, we perform a series of diagnostic analyses to understand how LLM-based GR models behave under standard random initialization and subsequent training. We focus our study on the embedding parameters of the injected SID tokens, which, due to the typical use of weight tying, govern both how the model interprets historical SIDs and how it generates recommendations. At a high level, we organize our analysis into three steps. First, we test the assumption that SIDs successfully impart semantics into the LLM by evaluating the semantic retention of the learned token embeddings. After discovering that this semantic information is largely discarded, we investigate what the embeddings actually encode, revealing a bias toward popularity. Finally, we provide a gradient analysis to explain why this bias emerges. 

\vspace{0.1cm}
\noindent \textbf{Datasets.} In this section, and the rest of the paper, we evaluate across five diverse sequential recommendation datasets including three domains from Amazon (Beauty, Sports, Toys), one from gaming (Steam), and one from movies (MovieLens). Details about the datasets are provided in \Cref{sec:hyperparams}.

\vspace{0.1cm}
\noindent \textbf{Experimental Setup.} We conduct our experiments using Qwen3-1.7B as the base LLM backbone \cite{yang2025qwen3technicalreport}, following recent LLM-based GR studies and reflecting the use of compact models for scalable GR \cite{chen2026llms, zhou2025openonerec, he2026reasoning, liu2025onerecthinkintextreasoninggenerative}. For each dataset, we expand the LLM vocabulary with SID tokens via standard random initialization (with mean and covariance of the base vocabulary) and evaluate the learned embeddings under two training paradigms: {pure-SFT} and {CPT+SFT}. Details on dataset processing and hyperparameters are provided in \Cref{sec:hyperparams}. For each diagnostic, we focus on the level-0 codebook $\mathcal{C}_0$ as, under residual quantization, losing semantic geometry at level 0 inherently prevents the LLM from preserving meaningful semantics in deeper levels.

\vspace{0.1cm}
\noindent \textit{Relation to Existing Pipelines.} We emphasize that the configurations evaluated in our experiments capture the common stages used by recent LLM-based GR systems. As prior work uses varying terminology for closely related stages, often bundling them with method-specific prompts or objectives, we label configurations by their underlying operations rather than individual method names. Our \emph{pure-SFT} setting captures vocabulary expansion followed directly by task fine-tuning, a standard paradigm used either directly or as a vanilla baseline in prior work \cite{penha2025semantic, zheng2024adapting, chen2026groundedtokeninitializationnew}.  Similarly, our \emph{CPT+SFT} setting captures the strategy of grounding SID tokens via language supervision prior to fine-tuning, leveraging the frozen-backbone optimization of GTI \cite{chen2026groundedtokeninitializationnew} and OneRec-Think \cite{liu2025onerecthinkintextreasoninggenerative}. Likewise, our prompting strategy mirrors OneRec-Think \cite{liu2025onerecthinkintextreasoninggenerative}, and closely aligns with PLUM's user behavior CPT prompts \cite{he2026plum}. This operation-based framing allows us to cleanly evaluate centroid initialization as a drop-in replacement across the foundational steps that govern modern LLM-based GR systems.

\subsection{Probing SID Embedding Geometry}
\label{sec:probing-geometry}

For our first analysis, we evaluate how well the LLM captures the categorical structure of the items as a measure of semantic information. Given that the premise of SIDs is that the discrete tokens reflect underlying item semantics, we expect well-trained SID embeddings within the LLM to exhibit local semantic coherence, e.g. two tokens with similar embeddings map to similar item categories. To test whether standard LLM-based GR training fulfills this expectation, we introduce a metric to quantify the semantic geometry of the learned tokens, detailed below.

\vspace{0.1cm}
\noindent \textbf{Local Semantic Neighborhoods.} To quantify how well the SID embeddings capture semantics, we define a neighborhood purity metric over the level-0 SID tokens. Let $\mathcal{N}_m(k)$ represent the $m$ nearest neighbors (via L2 distance) of code index $k\in\{0,\dots,K-1\}$ within the learned embedding space. We then compute the fraction of those neighbors that share the same ground-truth category as $k$:
\begin{equation}
P_m
=
\frac{1}{K}
\sum_{k=0}^{K-1}
\frac{1}{m}
\sum_{k'\in\mathcal{N}_m(k)}
\mathbb{1}\!\left[\mathrm{cat}_0(k')=\mathrm{cat}_0(k)\right].
\label{eq:purity}
\end{equation}
Intuitively, $P_m$ measures the semantic similarity of local regions in the embedding space. A high $P_m$ indicates that the model has organized the embeddings to reflect semantics, whereas a low $P_m$ suggests that the semantics were not encoded during training. 
 
\begin{figure}[t]
\centering
\includegraphics[width=0.48\textwidth]{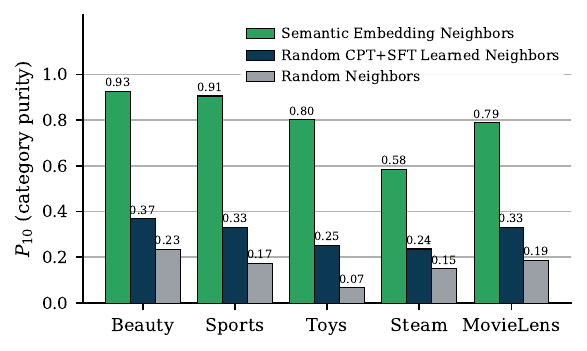}
\caption{Neighborhood purity ($P_{10}$) of learned SID embeddings under CPT+SFT, compared to a random baseline (gray) and the original semantic embeddings (green). While the semantic embeddings exhibit strong categorical coherence, standard random initialization fails to preserve this structure.}
\label{fig:deficiency}
\end{figure}

\vspace{0.1cm}
\noindent \textbf{Semantic Loss.} When evaluating the embeddings learned under random initialization with CPT+SFT, we find that they fail to maintain semantic purity. As illustrated in Figure~\ref{fig:deficiency}, across all five datasets, the 10-nearest neighbor purity ($P_{10}$) remains low, peaking at just 0.37 for Beauty and falling to 0.24 and 0.25 for Steam and Toys, respectively. Importantly, these values offer only a marginal improvement over drawing ten neighbors at random and computing $P_{10}$ (gray bars), and sit far below the purity in the original semantic embedding space (green bars). Thus, even after CPT, a stage designed to align SIDs with text and recover semantics, the embeddings do not possess this information. In the next section, we build on this finding to investigate what information the embeddings actually encode.

\subsection{Properties Encoded into SID Embeddings}
\label{sec:prop_enc}
Having established that the SID embeddings fail to retain semantic relationships, we now investigate what structural properties they acquire during training. Notably, the recommendation objective is driven by item-interaction sequences, providing a strong collaborative signal. Consequently, we hypothesize that without meaningful semantics, the LLM over-indexes on collaborative patterns, allowing behavioral artifacts (e.g., popularity) to dominate the recommendations. To test this hypothesis, we measure a series of complementary properties of the learned embeddings, detailed in the next sections.

\vspace{0.1cm}
\noindent \textbf{Spectral Scale.} First, we evaluate the capacity of the learned embedding space. Formally, let $\mathbf{X}\in\mathbb{R}^{K\times d_{\text{model}}}$ denote the centered embedding matrix for the level-0 SID tokens, with singular values $\sigma_1\ge\cdots\ge\sigma_r>0$. Then, we characterize the scale of the learned embeddings via:
\[
\sigma_{\mathrm{geo}}(\mathbf{X})
=
\left(\prod_{i=1}^{r}\sigma_i\right)^{1/r}.
\]
An increase in $\sigma_{\mathrm{geo}}$ signifies that the embedding space has expanded beyond the tightly compressed cluster created at initialization. 

\begin{table}[t]
\centering
\caption{
For the level-0 SID tokens, we report $\sigma_{\mathrm{geo}}$, $R_q^2(\mathbf{T}_{\mathrm{pop}})$, and $R_q^2(\mathbf{T}_{\mathrm{sem}})$. \textcolor[HTML]{5315f5}{Bold} denotes strong encoding ($> 0.75$). Standard initialization produces embeddings that
strongly encode popularity, discarding semantics even when CPT is applied. }
\label{tab:geom-random}
\setlength{\tabcolsep}{3.4pt}
\begin{tabular}{llc ccc ccc}
\toprule
& & & \multicolumn{3}{c}{$R_q^2(\mathbf{T}_{\mathrm{pop}})$}
    & \multicolumn{3}{c}{$R_q^2(\mathbf{T}_{\mathrm{sem}})$} \\
\cmidrule(lr){4-6}\cmidrule(lr){7-9}
Dataset & Training & $\sigma_{\mathrm{geo}}$
& $q{=}1$ & $q{=}3$ & $q{=}5$
& $q{=}1$ & $q{=}3$ & $q{=}5$ \\
\midrule
\multirow{2}{*}{Beauty}
    & pure-SFT & 0.16 & 0.00 & \textbf{\textcolor[HTML]{5315f5}{0.88}} & \textbf{\textcolor[HTML]{5315f5}{0.89}}
    & 0.01 & 0.15 & 0.25 \\
    & CPT+SFT  & 0.39 & 0.01 & \textbf{\textcolor[HTML]{5315f5}{0.84}} & \textbf{\textcolor[HTML]{5315f5}{0.86}}
    & 0.01 & 0.13 & 0.36 \\
\midrule
\multirow{2}{*}{Sports}
    & pure-SFT & 0.05 & 0.01 & \textbf{\textcolor[HTML]{5315f5}{0.80}} & \textbf{\textcolor[HTML]{5315f5}{0.82}}
    & 0.00 & 0.22 & 0.39 \\
    & CPT+SFT  & 0.44 & 0.00 & 0.22 & \textbf{\textcolor[HTML]{5315f5}{0.81}}
    & 0.00 & 0.22 & 0.35 \\
\midrule
\multirow{2}{*}{Toys}
    & pure-SFT & 0.04 & 0.00 & \textbf{\textcolor[HTML]{5315f5}{0.85}} & \textbf{\textcolor[HTML]{5315f5}{0.93}}
    & 0.01 & 0.05 & 0.13 \\
    & CPT+SFT  & 0.36 & 0.01 & 0.37 & \textbf{\textcolor[HTML]{5315f5}{0.77}}
    & 0.01 & 0.06 & 0.15 \\
\midrule
\multirow{2}{*}{Steam}
    & pure-SFT & 0.05 & 0.00 & \textbf{\textcolor[HTML]{5315f5}{0.87}} & \textbf{\textcolor[HTML]{5315f5}{0.91}}
    & 0.00 & 0.09 & 0.17 \\
    & CPT+SFT  & 0.43 & 0.01 & \textbf{\textcolor[HTML]{5315f5}{0.78}} & \textbf{\textcolor[HTML]{5315f5}{0.79}}
    & 0.00 & 0.07 & 0.11 \\
\midrule
\multirow{2}{*}{MovieLens}
    & pure-SFT & 0.06 & 0.00 & \textbf{\textcolor[HTML]{5315f5}{0.83}} & \textbf{\textcolor[HTML]{5315f5}{0.92}}
    & 0.00 & 0.02 & 0.12 \\
    & CPT+SFT  & 0.44 & 0.00 & \textbf{\textcolor[HTML]{5315f5}{0.78}} & \textbf{\textcolor[HTML]{5315f5}{0.80}}
    & 0.00 & 0.05 & 0.18 \\
\bottomrule
\end{tabular}
\end{table}

\vspace{0.1cm}
\noindent \textit{Relation to Previous Studies.} We explicitly adopt this scale-based metric to complement existing spectral analyses. While prior works have evaluated the spectral properties of SID embedding tables in LLMs, they typically focus on effective rank, demonstrating that the learned SID embeddings tend to collapse into low-dimensional subspaces \cite{chen2026groundedtokeninitializationnew}. However, effective rank is fundamentally scale-invariant, meaning it captures the distribution of variance but ignores the absolute spread of the embeddings. Consequently, effective rank can make random SID initializations appear structurally informative when, in reality, they merely form a small, compact ball of isotropic noise. In contrast, $\sigma_{\mathrm{geo}}$ reveals the true scale of the embedding space. 

\vspace{0.1cm}
\noindent \textbf{Latent Signal Encoding.} Next, we investigate which underlying signals govern the principal components (PCs) of the learned embedding space. Let $\mathbf{Z}_q\in\mathbb{R}^{K\times q}$ contain the top-$q$ left singular vectors of $\mathbf{X}$. We define two centered, per-code target matrices: $\mathbf{T}_{\mathrm{pop}}$ capturing code-level log popularity (aggregated over each code's item set), and $\mathbf{T}_{\mathrm{sem}}$ capturing the items' product category. Further details on how the targets are computed are provided in \Cref{sec:latent_signal_targets}. Then, for a given target $\mathbf{T} \in \{\mathbf{T}_{\mathrm{pop}}, \mathbf{T}_{\mathrm{sem}}\}$, we quantify its relationship with $\mathbf{Z}_q$ by computing the fraction of the target signal captured by the subspace spanned by $\mathbf{Z}_q$. This is expressed as:
\[
R_q^2(\mathbf{T})
=
\frac{\lVert \mathbf{Z}_q \mathbf{Z}_q^\top \mathbf{T}\rVert_F^2}{\lVert \mathbf{T}\rVert_F^2}.
\]

\vspace{0.1cm}
\noindent \textbf{Findings.} In Table~\ref{tab:geom-random}, we report $\sigma_{\mathrm{geo}}$ and the variance fractions $R_q^2(\mathbf{T})$ for $q \in \{1, 3, 5\}$. First, we find that embeddings learned under the standard random initialization are \textit{constrained in scale and largely dominated by popularity}. In the pure-SFT setting, all datasets remain tightly compressed ($\sigma_{\mathrm{geo}} \le 0.16$), and require only three PCs to capture the majority of the popularity signal (e.g., $R_3^2(\mathbf{T}_{\mathrm{pop}}) \ge 0.80$). When considering CPT+SFT, we find similar properties, where CPT only partially expands the embedding scale ($\sigma_{\mathrm{geo}} \approx 0.4$) and hardly impacts the dominance of popularity, as the top five PCs still heavily encode the popularity signal ($R_5^2(\mathbf{T}_{\mathrm{pop}}) \ge 0.77$).
Semantic encoding also remains low across both settings, where the top five PCs capture negligible categorical signal. This loss of semantics, even with CPT, exposes a fundamental issue with random initialization as the LLM over-indexes on heuristics like popularity. 

\subsection{How Training Can Encode Popularity}
\label{sec:analysis}
\label{sec:theory-main}

To provide intuition on how popularity is encoded, we analyze the gradient descent updates applied to the SID token embeddings. Then, by aggregating this update across multiple steps we explore a global approximation of the optimization behavior.

\vspace{0.1cm}
  \noindent\textbf{Setup.}
  Fix a level $\ell$ and consider the SID tokens associated with its codebook $\mathcal{C}_\ell$. For each code index
  $v\in\{0,\dots,K-1\}$, let $\mathbf{x}_v^{(t)}\in\R^{d_{\mathrm{model}}}$ denote its mean-centered output
  embedding at optimizer step $t$. Let
  $\widetilde n_v=n_v-\bar n$ denote its mean-centered supervision count, where
  $n_v$ is the number of times code $v$ appears as a target and $\bar n$ is the
  average target count across the level-$\ell$ codebook. Finally, let $\bar{\mathbf{h}}^{(t)}\in\R^{d_{\mathrm{model}}}$ denote the average final-layer hidden state used to predict any target token associated with $\mathcal{C}_\ell$, taken across all sequences in the training set.

 \begin{theorem}[Popularity Accumulation in SID Embeddings]
  \label{thm:accumulated-bias-main}
   Suppose an SID embedding is trained for $S$ steps using
  full-batch gradient descent with fixed step size $\eta$ on the summed cross-entropy loss. Then, for a code
  $v\in\{0,\dots,K-1\}$, its mean-centered embedding is given by
    \begin{equation}
      \mathbf{x}_v^{(S)} = \mathbf{x}_v^{(0)} + \eta\widetilde n_v \sum_{t=0}^{S-1}\bar{\mathbf{h}}^{(t)} + \eta\sum_{t=0}^{S-1}\boldsymbol{\Delta}_v^{(t)},
      \label{eq:accumulated-main}
  \end{equation}
  where the vector $\boldsymbol{\Delta}_v^{(t)} \in \R^{d_{\mathrm{model}}}$ is a term capturing the context- and model-dependent gradient dynamics specific to code $v$.
  \end{theorem}

\noindent 
    The proof and $\boldsymbol{\Delta}_v^{(t)}$ details are provided in \Cref{sec:theory_app}. \hfill $\square$

\vspace{0.1cm}
\noindent\textbf{Prediction-Time Implication.}
The theorem above characterizes a \textit{training-time} accumulation mechanism. Let $\mathbf{s}=\sum_{t=0}^{S-1}\bar{\mathbf{h}}^{(t)}$ denote the sum of the average final hidden states. Then, the term $\eta\widetilde n_v \mathbf{s}$ encodes a popularity-aligned axis into the embedding space. Geometrically, this component shifts every code along the shared direction $\mathbf{s}$ by a distance dictated by its relative popularity. Codes appearing more often than average ($\widetilde n_v>0$) are pushed forward along $+\mathbf{s}$, while rarer codes ($\widetilde n_v<0$) are pushed in the opposite direction. 

To understand the implications of this behavior, consider the final-layer hidden state $\mathbf{h}$ produced by the trained LLM at \textit{prediction-time}. As the LLM computes logits via inner products with $\mathbf{h}$, this operation can effectively {read out} the popularity information encoded during training. Specifically, for the codebook-centered logit $\widetilde z_v(\mathbf{h})=\langle\mathbf{h},\mathbf{x}_v^{(S)}\rangle$, the frequency-driven contribution is: 
\[ 
    \delta z_{v,\mathrm{freq}}(\mathbf{h}) = \eta\widetilde n_v \left\langle\mathbf{h},\mathbf{s}\right\rangle. 
\] 
Then, the relative score between two codes $u$ and $v$ is: 
\[ 
    \delta\!\left(z_u(\mathbf{h})-z_v(\mathbf{h})\right)_{\mathrm{freq}} = \eta(n_u-n_v) \left\langle\mathbf{h},\mathbf{s}\right\rangle. 
\] 
Thus, if code $u$ is more popular than $v$ ($n_u>n_v$), $\delta z_{u,\mathrm{freq}}(\mathbf{h})$ can boost the logit of $u$ relative to $v$ whenever $\langle\mathbf{h},\mathbf{s}\rangle>0$. Importantly, while the final exact logits depend on the interplay of all three terms in \Cref{thm:accumulated-bias-main}, this decomposition isolates a distinct, training-induced component that structurally favors popular SID codes. We further discuss the scope of this analysis in \Cref{sec:theory_app}.

\section{Preserving Semantics via SID Centroids}
\label{sec:centroids}

In Section~\ref{sec:motivation}, we showed that standard LLM-based GR pipelines fail to preserve the underlying semantics of the items, encoding popularity instead. To address this issue, we consider how one can directly imbue the expanded vocabulary with the embeddings used to derive the SIDs, minimizing the loss of semantics. Our philosophy for this intervention centers on avoiding changes to the training objectives, model architecture, or inference procedure, for easy adoption. Thus, by injecting this prior, we provide the LLM with a rich geometric structure,  \textit{preserving item semantics for free}.

\begin{algorithm}[t]
  \caption{\textbf{Pseudocode for Centroid Init.}
  We mean-shift RQ centroids into the pretrained token space and map them to SID rows.}
  \label{alg:remap}

\begin{lstlisting}[style=pytorch-muted,firstnumber=1,]
# C: RQ centroids [num_codes, hidden_dim]
# sid_ids: SID token indices in vocab (not de-dup tokens)
# V: number of pre-existing, non-SID vocabulary rows

# Standard SID vocabulary expansion
model.resize_token_embeddings(len(tok), mean_resizing=True)
\end{lstlisting}
\begin{lstlisting}[style=pytorch-focus,firstnumber=last]

# (Ours) Centroid initialization with Matryoshka
E = model.get_input_embeddings().weight
C = torch.load("centroids.pt").to(E)
C = C - C.mean(0) + E[:V].mean(0)
with torch.no_grad(): E[sid_ids] = C
\end{lstlisting}

\begin{lstlisting}[style=pytorch-muted,firstnumber=last]

# Existing pure-SFT or CPT+SFT training is unchanged
for history, target_sid in dataloader:
   ...
\end{lstlisting}
\end{algorithm}
\subsection{Mean-Shifted Centroid Initialization}
\label{sec:method-init}
Geometrically, an SID token represents a volume in the semantic embedding space, defined by the boundaries induced by the quantizer. Consequently, the ideal proxy for this volume is its most representative point, i.e., its centroid. Therefore, we focus on leveraging these centroids to create a bridge between the original semantic space and the LLM's embedding space. 
To minimize overhead, we propose a straightforward intervention of \textit{initializing the SID token embeddings with their corresponding centroids},  providing the LLM with a semantic prior for training.
Formally, under residual $k$-means, let SID code index $k$ at level $\ell$ correspond to a centroid $\mathbf{c}_{\ell,k} \in \mathbb{R}^{d}$ in the semantic embedding space. Additionally, assume the LLM has embedding matrix $\mathbf{E}\in\mathbb{R}^{|\mathcal{V}|\times d_{\text{model}}}$. To effectively bridge these two spaces, our initialization strategy is guided by two core design principles:
\begin{enumerate}[leftmargin=*]
\item \textbf{Dimension Matching:} As $d$ and $d_{\text{model}}$ typically differ, we apply a mapping from the text embedding space to the LLM space.
\item \textbf{Optimization Stability:} To inherit the stable training of standard random initialization, we apply a mean-shift translation that aligns the centroids with the pre-trained token distribution.
\end{enumerate}
Combining these, we mathematically define the initialized embedding $\mathbf{e}^{\mathrm{cent}}_{\ell,k}$ for each semantic code $k$ at level $\ell \in \{0,\dots,L-1\}$ as:
\begin{equation}
\mathbf{e}^{\mathrm{cent}}_{\ell,k}
=
\Phi_{d_{\text{model}}}(\mathbf{c}_{\ell,k} - \bar{\mathbf{c}}) + \boldsymbol{\mu}_{\mathrm{base}},
\label{eq:centinit}
\end{equation}
where $\bar{\mathbf{c}}$ is the mean across all semantic centroids, $\boldsymbol{\mu}_{\mathrm{base}}$ is the mean of the original LLM token embeddings, and $\Phi_{d_{\text{model}}}(\cdot)$ represents a function that projects a vector down to $d_{\text{model}}$.

While $\Phi$ can theoretically be any learned projection, introducing parametric transformations risks distorting the original semantic geometry. To bypass this, our key insight is to exploit the Matryoshka Representation Learning (MRL) property in modern text embeddings \cite{kusupati2022matryoshka, nussbaumnomic, li2026qwen3}. As MRL places critical information into the leading dimensions, we define $\Phi_{d_{\text{model}}}(\cdot)$ as a truncation to the first $d_{\text{model}}$ dimensions, retaining the embeddings' semantics with zero computational overhead and no additional parameters. This truncation also informs our choice of standard residual $k$-means over parametric quantizers like RQ-VAE. While RQ-VAE projects items into a learned latent space, $k$-means operates directly within the native semantic embedding space, allowing us to exploit MRL. For deduplication tokens, which do not correspond to any clusters, we use random initialization. Pseudocode can be found in \Cref{alg:remap}.

\begin{figure}[t]
\centering
\includegraphics[width=0.48\textwidth]{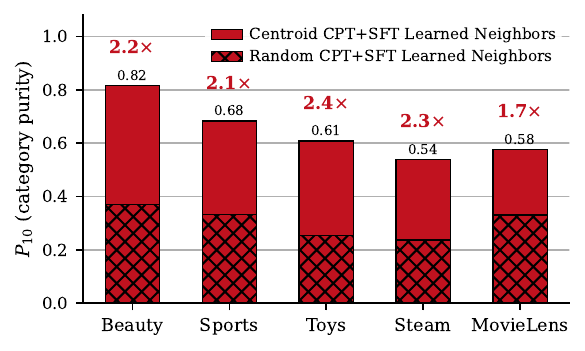}
\caption{Neighborhood purity ($P_{10}$) of SID embeddings under centroid initialization and CPT+SFT, vs. random initialization. Values above bars indicate the increase in purity. Centroid initialization enhances the semantic structure of the latent space.}
\label{fig:deficiency_cent}
\end{figure}
\subsection{\textit{Re}-Probing SID Embeddings}
\label{sec:method-geometry}

To verify that the concerns in Section~\ref{sec:motivation} are resolved, we re-perform the diagnostic analyses after training with centroid initialization.

\begin{table}[b]
\centering
\caption{For the level-0 SID tokens, we report the change ($\Delta$) in $\sigma_{\mathrm{geo}}$, $R_q^2(\mathbf{T}_{\mathrm{pop}})$, and $R_q^2(\mathbf{T}_{\mathrm{sem}})$ relative to random initialization (Table~\ref{tab:geom-random}). Desired outcomes (increases in scale/semantics, decreases in popularity) are colored green, while undesired outcomes (increases in popularity) are colored red. Color is scaled to the maximum observed value in $\mathbf{T}_{\mathrm{pop}}$ and $\mathbf{T}_{\mathrm{sem}}$.}
\label{tab:geom-centroid}
\setlength{\tabcolsep}{1.8pt}
\begin{tabular}{llc ccc ccc}
\toprule
& & & \multicolumn{3}{c}{$\Delta R_q^2(\mathbf{T}_{\mathrm{pop}})$}
    & \multicolumn{3}{c}{$\Delta R_q^2(\mathbf{T}_{\mathrm{sem}})$} \\
\cmidrule(lr){4-6}\cmidrule(lr){7-9}
Dataset & Stage & $\Delta \sigma_{\mathrm{geo}}$
& $q{=}1$ & $q{=}3$ & $q{=}5$
& $q{=}1$ & $q{=}3$ & $q{=}5$ \\
\midrule
\multirow{2}{*}{Beauty}
    & pure-SFT & \textcolor{green!31!black}{+0.13} & 0.00 & \textcolor{green!98!black}{-0.88} & \textcolor{green!92!black}{-0.83}
    & \textcolor{green!43!black}{+0.18} & \textcolor{green!88!black}{+0.37} & \textcolor{green!93!black}{+0.39} \\
    & CPT+SFT  & \textcolor{green!33!black}{+0.14} & \textcolor{green!1!black}{-0.01} & \textcolor{green!92!black}{-0.83} & \textcolor{green!91!black}{-0.82}
    & \textcolor{green!43!black}{+0.18} & \textcolor{green!100!black}{+0.42} & \textcolor{green!69!black}{+0.29} \\
\midrule
\multirow{2}{*}{Sports}
    & pure-SFT & \textcolor{green!50!black}{+0.21} & 0.00 & \textcolor{green!82!black}{-0.74} & \textcolor{green!78!black}{-0.70}
    & \textcolor{green!55!black}{+0.23} & \textcolor{green!45!black}{+0.19} & \textcolor{green!17!black}{+0.07} \\
    & CPT+SFT  & \textcolor{green!31!black}{+0.13} & \textcolor{red!5!black}{+0.02} & \textcolor{green!12!black}{-0.11} & \textcolor{green!69!black}{-0.62}
    & \textcolor{green!55!black}{+0.23} & \textcolor{green!50!black}{+0.21} & \textcolor{green!29!black}{+0.12} \\
\midrule
\multirow{2}{*}{Toys}
    & pure-SFT & \textcolor{green!50!black}{+0.21} & 0.00 & \textcolor{green!92!black}{-0.83} & \textcolor{green!100!black}{-0.90}
    & \textcolor{green!10!black}{+0.04} & \textcolor{green!31!black}{+0.13} & \textcolor{green!33!black}{+0.14} \\
    & CPT+SFT  & \textcolor{green!31!black}{+0.13} & 0.00 & \textcolor{green!38!black}{-0.34} & \textcolor{green!81!black}{-0.73}
    & \textcolor{green!10!black}{+0.04} & \textcolor{green!29!black}{+0.12} & \textcolor{green!31!black}{+0.13} \\
\midrule
\multirow{2}{*}{Steam}
    & pure-SFT & \textcolor{green!33!black}{+0.14} & \textcolor{red!29!black}{+0.12} & \textcolor{green!66!black}{-0.59} & \textcolor{green!69!black}{-0.62}
    & \textcolor{green!21!black}{+0.09} & \textcolor{green!50!black}{+0.21} & \textcolor{green!43!black}{+0.18} \\
    & CPT+SFT  & \textcolor{green!21!black}{+0.09} & \textcolor{red!79!black}{+0.33} & \textcolor{green!24!black}{-0.22} & \textcolor{green!24!black}{-0.22}
    & \textcolor{green!21!black}{+0.09} & \textcolor{green!50!black}{+0.21} & \textcolor{green!52!black}{+0.22} \\
\midrule
\multirow{2}{*}{MovieLens}
    & pure-SFT & \textcolor{green!21!black}{+0.09} & \textcolor{red!7!black}{+0.03} & \textcolor{green!81!black}{-0.73} & \textcolor{green!59!black}{-0.53}
    & \textcolor{green!50!black}{+0.21} & \textcolor{green!90!black}{+0.38} & \textcolor{green!100!black}{+0.42} \\
    & CPT+SFT  & \textcolor{green!14!black}{+0.06} & \textcolor{red!5!black}{+0.02} & \textcolor{green!60!black}{-0.54} & \textcolor{green!11!black}{-0.10}
    & \textcolor{green!7!black}{+0.03} & \textcolor{green!74!black}{+0.31} & \textcolor{green!74!black}{+0.31} \\
\bottomrule
\end{tabular}
\end{table}

\begin{figure*}[t]
\centering
\includegraphics[width=0.94\textwidth]{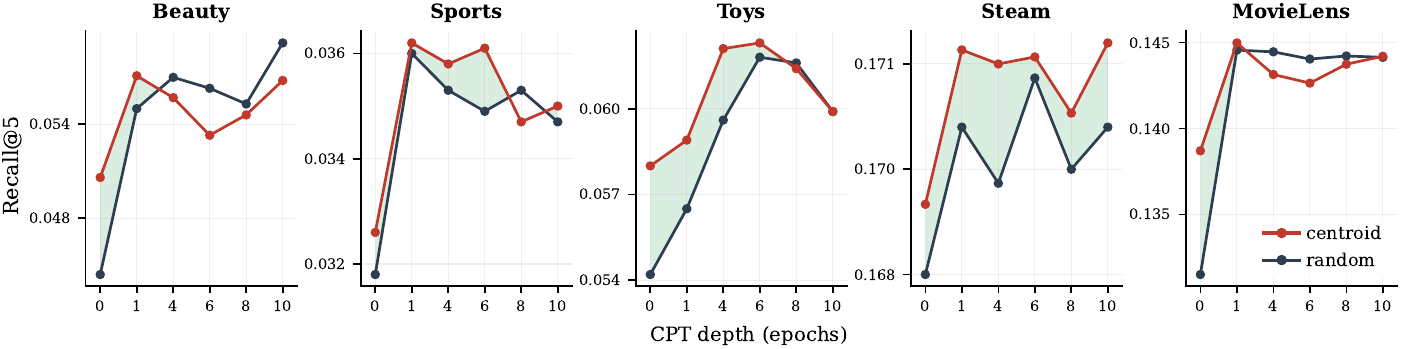}
\caption{
Recall@5 across varying CPT depths. Centroid-init can reach the random-init performance with fewer CPT epochs.}
\label{fig:cptdepth}
\end{figure*}

\vspace{0.1cm}
\noindent \textbf{Recovering Semantics.} First, we evaluate whether the learned embeddings approach the category purity of the semantic embeddings seen in Figure~\ref{fig:deficiency}. As illustrated in Figure~\ref{fig:deficiency_cent}, we find that centroid initialization significantly improves the neighbor purity ($P_{10}$) of the learned representations. Compared to random initialization, where embeddings barely outperformed random chance even after CPT, directly injecting the semantic prior allows the embeddings to form meaningful categorical clusters, often doubling the baseline purity.

\vspace{0.1cm}
\noindent \textbf{Restoring Scale.} Next, we examine the scale occupied by the level-0 SID tokens under centroid initialization. As shown in Table~\ref{tab:geom-centroid}, for pure-SFT, the embedding scale expands significantly, e.g., $\sigma_{\mathrm{geo}}$ increases from $0.16$ with random initialization to $0.29$ on Beauty. When combined with CPT, the space expands even further to $\sigma_{\mathrm{geo}} > 0.50$ across all datasets, signifying better utilization.

\vspace{0.1cm}
\noindent \textbf{Overcoming Popularity.} In Table~\ref{tab:geom-centroid}, we show centroid initialization dampens the encoding of popularity in the learned embeddings. Across all datasets, the popularity variance captured by the leading PCs
drops drastically, e.g., $R_5^2(\mathbf{T}_{\mathrm{pop}})$ goes from $0.89$ to $0.06$ for Beauty trained with pure-SFT. In its place, the leading PCs shift to encoding item category information, with the captured semantic variance ($R_5^2(\mathbf{T}_{\mathrm{sem}})$) increasing to $0.64$ for Beauty and $0.54$ for MovieLens. Together, these results show that centroid initialization enables the LLM to better retain item semantics rather than encoding popularity.

\section{Practical Benefits of Centroid Initialization}
\label{sec:empirical-benefits}

\begin{table}[b]
\centering
\caption{\textbf{Centroid initialization improves pure-SFT and reaches its peak earlier.} Here, $t^\star$ is the optimizer step with the highest validation Recall@5. \textcolor[HTML]{5315f5}{\textbf{Bold}} is better.}
\label{tab:av1}
\begin{tabular}{ll ccc}
\toprule
Domain & Init. & Recall@5 & Recall@10 & peak step $t^\star\!\downarrow$ \\
\midrule
\multirow{2}{*}{Beauty}
& Random   & 0.044 & 0.067 & 6989 \\
& Centroid & \textbf{\textcolor[HTML]{5315f5}{0.051}} & \textbf{\textcolor[HTML]{5315f5}{0.073}} & \textbf{\textcolor[HTML]{5315f5}{4193}} \\
\midrule
\multirow{2}{*}{Sports}
& Random   & 0.032 & 0.046 & 17799 \\
& Centroid & \textbf{\textcolor[HTML]{5315f5}{0.033}} & \textbf{\textcolor[HTML]{5315f5}{0.047}} & \textbf{\textcolor[HTML]{5315f5}{13349}} \\
\midrule
\multirow{2}{*}{Toys}
& Random   & 0.054 & 0.075 & 10925 \\
& Centroid & \textbf{\textcolor[HTML]{5315f5}{0.058}} & \textbf{\textcolor[HTML]{5315f5}{0.081}} & \textbf{\textcolor[HTML]{5315f5}{7283}} \\
\midrule
\multirow{2}{*}{Steam}
& Random   & 0.168 & \textbf{\textcolor[HTML]{5315f5}{0.198}} & 14930 \\
& Centroid & \textbf{\textcolor[HTML]{5315f5}{0.169}} & \textbf{\textcolor[HTML]{5315f5}{0.198}} & \textbf{\textcolor[HTML]{5315f5}{11940}} \\
\midrule
\multirow{2}{*}{MovieLens}
& Random   & 0.132 & 0.177 & 18749 \\
& Centroid & \textbf{\textcolor[HTML]{5315f5}{0.139}} & \textbf{\textcolor[HTML]{5315f5}{0.187}} & \textbf{\textcolor[HTML]{5315f5}{15624}} \\
\bottomrule
\end{tabular}
\end{table}

Having established that centroid initialization successfully encourages the encoding of semantics, we now verify that this new geometry supports recommendation. We focus on three core areas: 

\begin{itemize}[leftmargin=*]
    \item \textbf{Pure-SFT:} We show that centroid initialization can compensate for the lack of CPT, enabling {recall gains and faster convergence}.
    
    \item \textbf{CPT+SFT:} We show that centroid initialization can help {reduce the number of CPT epochs needed to attain peak performance}.
    
    \item \textbf{Generalization:} We demonstrate {improvements for recall on cold items} and benefits across {backbone size scaling}.
\end{itemize}

\subsection{Improving LLM-based GR Training}
\label{sec:acceleration}

In this section, we evaluate centroid initialization on recommendation quality and overall training efficiency across LLM-based GR pipelines. The experimental setup matches \Cref{sec:prop_enc}, except for the scaling study where the Qwen3 backbone size varies.

\vspace{0.1cm}
\noindent \textbf{Pure-SFT: A Semantic Warm Start.} We first analyze the pure-SFT setting to cleanly isolate the impact of centroid initialization while also reflecting compute-limited environments where only SFT is feasible.
As shown in Table~\ref{tab:av1}, centroid initialization significantly improves over random initialization, producing increases in recall and training efficiency. Across all datasets, centroid initialization yields an average relative improvement of 6.5\% in Recall@5 while simultaneously reducing the time-to-convergence by an average of 27\%. More specifically, Beauty and Toys see Recall@5 improvements of 15.9\% and 7.4\%, while requiring 40\% and 33\% fewer training steps, respectively.
We attribute these gains to the geometric warm start — as centroid initialization directly encodes semantics, the optimizer can skip this step and immediately learn from historical item interactions. Additional NDCG results can be found in \Cref{tab:av1-ndcg}.

\vspace{0.1cm}
\noindent \textbf{CPT+SFT: Expediting Convergence.} We next evaluate the full CPT+SFT pipeline, where the model undergoes semantic alignment prior to SFT. Since CPT enables the random baseline to partially encode semantic information, absolute accuracy gains are less pronounced than in the pure SFT regime. However, our intervention introduces a separate, useful benefit by reducing the number of CPT epochs required to achieve comparable performance. As illustrated in \Cref{fig:cptdepth}, the centroid-initialized models reach their optimal Recall@5 in a fraction of the time required by the standard initialization. On average across the five datasets, a centroid-initialized model with just 1 to 4 epochs of CPT can match or exceed the peak performance of a randomly initialized model trained for 6 to 10 epochs. Beauty is the one exception where the random baseline achieves slightly higher performance, though this requires $10\times$ as many CPT epochs. We attribute this reduction directly to the geometric warm start, which natively encodes item semantics so the CPT phase only has to align the pre-existing structure with the LLM text space rather than rebuilding it. Additional Recall@10 and NDCG results can be found in \Cref{fig:cptdepth_app}.

\subsection{Centroids Improve Cold Item Generalization}
\label{sec:tail-cold}

To better understand \textit{how} centroid initialization drives improvement, we focus on its ability to decouple item similarity from item-interaction history. Fundamentally, semantic priors allow the model to recognize related items, such as two products from the same niche hobby, without needing to observe direct co-purchases. Consequently, we hypothesize that centroid initialization should offer the most benefit for items with sparse interaction histories, where collaborative signals are weak and semantics are the only reliable proxy. To test this hypothesis, we isolate the performance on cold items by measuring a cohort-specific recall, focusing strictly on users whose target test items appear at most once as a supervision target in the training set. 

\begin{figure}[t]
\centering
\includegraphics[width=0.48\textwidth]{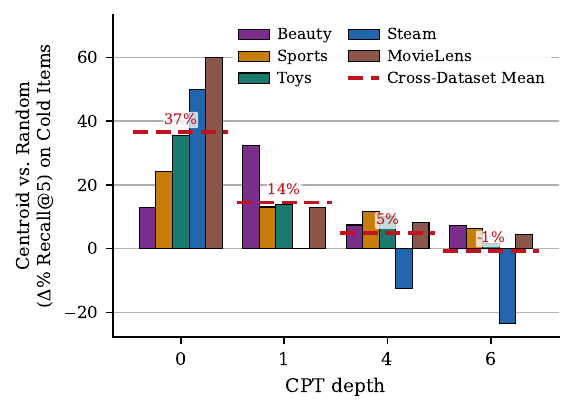}
\caption{Relative Recall@5 gap (centroid $-$ random) on cold targets (train-target freq.\ $\leq\!1$) versus the number of CPT epochs. With fewer CPT epochs, centroid initialization tends to produce larger cold-item recall gains.}
\label{fig:tierdepth_cold}
\end{figure}
In \Cref{fig:tierdepth_cold}, we demonstrate that this intuition holds, with centroid initialization producing large performance gains for users with cold item targets. In the pure-SFT regime (zero CPT epochs), we find that centroid initialization improves Recall@5 on cold items, on average, by 37\% compared to the random baseline. Specifically, on MovieLens and Steam, these gains reach 60\% and 50\%, respectively. Additionally, even with one epoch of CPT, the centroid-initialized models maintain a strong 14\% cross-dataset advantage over standard initialization.
Beyond performance, this behavior also offers fundamental insight into the dynamics of CPT+SFT training. As seen in \Cref{sec:cptdebias}, increasing the number of CPT epochs steadily improves cold-item performance even for randomly initialized models. Therefore, while the relative advantage of centroid initialization naturally diminishes at later epochs (e.g., 4 and 6), this convergence explicitly illustrates \textit{how} our intervention saves compute. Rather than forcing the model to slowly learn how to align and recommend cold items across multiple CPT epochs, centroid initialization provides that information upfront. Additional Recall@10 and NDCG results can be found in \Cref{fig:tierdepth_cold_app}.

\vspace{0.1cm}
\noindent \textbf{Robustness Across Scale.} To investigate whether benefits persist across LLM backbone scales, we evaluate 0.6B and 4B Qwen3 models on the Amazon domains (\Cref{tab:backbones}). At 0.6B, centroid initialization consistently outperforms the random baseline across all regimes, suggesting the semantic prior provides value not fully recovered by CPT in smaller models. Furthermore, it significantly accelerates pure-SFT convergence, peaking 40\% and 25\% earlier on Toys and Sports, respectively, and reaching the baseline's peak 29\% earlier on Beauty. At the 4B scale, these convergence benefits further increase, with the model peaking 43\%, 40\%, and 17\% earlier on Toys, Sports, and Beauty. In terms of 4B performance, centroid initialization improves pure-SFT Recall@5 on Toys (+9.8\%) and Sports (+2.9\%). While the random baseline attains marginally higher pure-SFT Recall@5 on Beauty, mirroring \Cref{fig:cptdepth}, this edge requires roughly 20\% more optimizer steps than the centroid model's peak. Under the CPT+SFT regime at the 4B scale, centroid initialization provides small but consistent improvements to Recall@5 across all three domains, alongside Recall@10 improvements on Beauty and Sports. Together, these results demonstrate that the semantic prior delivers a robust benefit across model scales, where recall gains remain prominent at lower capacities while convergence acceleration scales effectively with model size.

\begin{table}[]
\centering
\caption{\textbf{Scale study of Qwen3.} We evaluate random vs. centroid initialization across pure-SFT and CPT+SFT at the 0.6B and 4B scales, reporting Recall@5 and Recall@10 and highlighting centroid improvements in \textcolor[HTML]{5315f5}{\textbf{bold}}.}
\label{tab:backbones}
\setlength{\tabcolsep}{2pt}
\begin{tabular}{lllcccccc}
\toprule
 & Training & Init & \multicolumn{2}{c}{Beauty} & \multicolumn{2}{c}{Sports} & \multicolumn{2}{c}{Toys} \\
\cmidrule(lr){4-5}\cmidrule(lr){6-7}\cmidrule(lr){8-9}
 & & & R@5 & R@10 & R@5 & R@10 & R@5 & R@10 \\
\midrule
\multirow{4}{*}{0.6B}
 & \multirow{2}{*}{Pure-SFT} & Random   & 0.047 & 0.068 & 0.032 & 0.047 & 0.051 & 0.073 \\
 &                           & Centroid & \textbf{\textcolor[HTML]{5315f5}{0.051}} & \textbf{\textcolor[HTML]{5315f5}{0.074}} & \textbf{\textcolor[HTML]{5315f5}{0.033}} & \textbf{\textcolor[HTML]{5315f5}{0.047}} & \textbf{\textcolor[HTML]{5315f5}{0.056}} & \textbf{\textcolor[HTML]{5315f5}{0.078}} \\
\cmidrule(lr){2-9}
 & \multirow{2}{*}{CPT+SFT}  & Random   & 0.054 & 0.077 & 0.033 & 0.048 & 0.056 & 0.080 \\
 &                           & Centroid & \textbf{\textcolor[HTML]{5315f5}{0.056}} & \textbf{\textcolor[HTML]{5315f5}{0.078}} & \textbf{\textcolor[HTML]{5315f5}{0.034}} & \textbf{\textcolor[HTML]{5315f5}{0.049}} & \textbf{\textcolor[HTML]{5315f5}{0.058}} & \textbf{\textcolor[HTML]{5315f5}{0.083}} \\
\midrule
\multirow{4}{*}{4B}
 & \multirow{2}{*}{Pure-SFT} & Random   & 0.051 & 0.075 & 0.033 & 0.047 & 0.053 & 0.077 \\
 &                           & Centroid & 0.050 & 0.072 & \textbf{\textcolor[HTML]{5315f5}{0.034}} & 0.047 & \textbf{\textcolor[HTML]{5315f5}{0.058}} & \textbf{\textcolor[HTML]{5315f5}{0.082}} \\
\cmidrule(lr){2-9}
 & \multirow{2}{*}{CPT+SFT}  & Random   & 0.056 & 0.079 & 0.033 & 0.048 & 0.060 & 0.086 \\
 &                           & Centroid & \textbf{\textcolor[HTML]{5315f5}{0.058}} & \textbf{\textcolor[HTML]{5315f5}{0.082}} & \textbf{\textcolor[HTML]{5315f5}{0.035}} & \textbf{\textcolor[HTML]{5315f5}{0.049}} & \textbf{\textcolor[HTML]{5315f5}{0.061}} & 0.085 \\
\bottomrule

\end{tabular}
\end{table}

\section{Conclusion}
In this work, we investigated the behavior of LLM-based GR models, seeking to uncover what SID embeddings learn during training. Through comprehensive analyses, we found that the practice of using randomly initialized SID token embeddings failed to recover the original semantics of the item corpus, regressing instead to the well-known biases of traditional recommenders. To overcome this, we introduced a simple yet effective strategy: directly injecting the centroids of the original SID embedding space into the LLM's token space. Empirically, this unlocked several beneficial properties, including (i) accelerated training convergence, (ii) a reduced need for explicit pre-training, and (iii) recall improvements, particularly for cold items. Given our findings, we believe this semantically informed initialization moves the field closer to the true promise of LLM-based GR, enabling better utilization of both collaborative and semantic signals.

\bibliographystyle{ACM-Reference-Format}
\bibliography{references}

@article{gomez2016netflix,
author = {Gomez-Uribe, Carlos A. and Hunt, Neil},
title = {The Netflix Recommender System: Algorithms, Business Value, and Innovation},
year = {2016},
issue_date = {January 2016},
publisher = {Association for Computing Machinery},
address = {New York, NY, USA},
volume = {6},
number = {4},
issn = {2158-656X},
url = {https://doi.org/10.1145/2843948},
doi = {10.1145/2843948},
journal = {ACM Trans. Manage. Inf. Syst.},
month = dec,
articleno = {13},
numpages = {19}
}

@inproceedings{liu2019recsyspersonal,
author = {Liu, Huafeng and Wen, Jingxuan and Jing, Liping and Yu, Jian},
title = {Deep generative ranking for personalized recommendation},
year = {2019},
isbn = {9781450362436},
publisher = {Association for Computing Machinery},
address = {New York, NY, USA},
url = {https://doi.org/10.1145/3298689.3347012},
booktitle = {Proceedings of the 13th ACM Conference on Recommender Systems},
pages = {34–42},
numpages = {9},
location = {Copenhagen, Denmark},
series = {RecSys '19}
}

@article{rajput2023recommender,
  title={Recommender systems with generative retrieval},
  author={Rajput, Shashank and Mehta, Nikhil and Singh, Anima and Hulikal Keshavan, Raghunandan and Vu, Trung and Heldt, Lukasz and Hong, Lichan and Tay, Yi and Tran, Vinh and Samost, Jonah and others},
  journal={Advances in Neural Information Processing Systems},
  volume={36},
  pages={10299--10315},
  year={2023}
}

@inproceedings{fan2019graphrec,
author = {Fan, Wenqi and Ma, Yao and Li, Qing and He, Yuan and Zhao, Eric and Tang, Jiliang and Yin, Dawei},
title = {Graph Neural Networks for Social Recommendation},
year = {2019},
isbn = {9781450366748},
publisher = {Association for Computing Machinery},
address = {New York, NY, USA},
url = {https://doi.org/10.1145/3308558.3313488},
doi = {10.1145/3308558.3313488},
booktitle = {The World Wide Web Conference},
pages = {417–426},
numpages = {10},
location = {San Francisco, CA, USA},
series = {WWW '19}
}

@inproceedings{schafer1999recec,
author = {Schafer, J. Ben and Konstan, Joseph and Riedl, John},
title = {Recommender systems in e-commerce},
year = {1999},
isbn = {1581131763},
publisher = {Association for Computing Machinery},
address = {New York, NY, USA},
url = {https://doi.org/10.1145/336992.337035},
doi = {10.1145/336992.337035},
booktitle = {Proceedings of the 1st ACM Conference on Electronic Commerce},
pages = {158–166},
numpages = {9},
location = {Denver, Colorado, USA},
series = {EC '99}
}

@article{jin2024amazon,
  title={Amazon-m2: A multilingual multi-locale shopping session dataset for recommendation and text generation},
  author={Jin, Wei and Mao, Haitao and Li, Zheng and Jiang, Haoming and Luo, Chen and Wen, Hongzhi and Han, Haoyu and Lu, Hanqing and Wang, Zhengyang and Li, Ruirui and others},
  journal={Advances in Neural Information Processing Systems},
  volume={36},
  year={2024}
}

@inproceedings{wang2024eager,
  title={Eager: Two-stream generative recommender with behavior-semantic collaboration},
  author={Wang, Ye and Xun, Jiahao and Hong, Minjie and Zhu, Jieming and Jin, Tao and Lin, Wang and Li, Haoyuan and Li, Linjun and Xia, Yan and Zhao, Zhou and others},
  booktitle={Proceedings of the 30th ACM SIGKDD Conference on Knowledge Discovery and Data Mining},
  pages={3245--3254},
  year={2024}
}

@inproceedings{zhai2024actions,
  title={Actions Speak Louder than Words: Trillion-Parameter Sequential Transducers for Generative Recommendations},
  author={Zhai, Jiaqi and Liao, Lucy and Liu, Xing and Wang, Yueming and Li, Rui and Cao, Xuan and Gao, Leon and Gong, Zhaojie and Gu, Fangda and He, Jiayuan and others},
  booktitle={International Conference on Machine Learning},
  pages={58484--58509},
  year={2024},
  organization={PMLR}
}

@misc{deng2025onerecunifyingretrieverank,
      title={OneRec: Unifying Retrieve and Rank with Generative Recommender and Iterative Preference Alignment}, 
      author={Jiaxin Deng and Shiyao Wang and Kuo Cai and Lejian Ren and Qigen Hu and Weifeng Ding and Qiang Luo and Guorui Zhou},
      year={2025},
      eprint={2502.18965},
      archivePrefix={arXiv},
      primaryClass={cs.IR},
      url={https://arxiv.org/abs/2502.18965}, 
}

@inproceedings{hou2024large,
  title={Large language models are zero-shot rankers for recommender systems},
  author={Hou, Yupeng and Zhang, Junjie and Lin, Zihan and Lu, Hongyu and Xie, Ruobing and McAuley, Julian and Zhao, Wayne Xin},
  booktitle={European conference on information retrieval},
  pages={364--381},
  year={2024},
  organization={Springer}
}

@misc{MiniOneRec,
      title={MiniOneRec: An Open-Source Framework for Scaling Generative Recommendation}, 
      author={Xiaoyu Kong and Leheng Sheng and Junfei Tan and Yuxin Chen and Jiancan Wu and An Zhang and Xiang Wang and Xiangnan He},
      year={2025},
      eprint={2510.24431},
      archivePrefix={arXiv},
      primaryClass={cs.IR},
}

@misc{liu2025onerecthinkintextreasoninggenerative,
      title={OneRec-Think: In-Text Reasoning for Generative Recommendation}, 
      author={Zhanyu Liu and Shiyao Wang and Xingmei Wang and Rongzhou Zhang and Jiaxin Deng and Honghui Bao and Jinghao Zhang and Wuchao Li and Pengfei Zheng and Xiangyu Wu and Yifei Hu and Qigen Hu and Xinchen Luo and Lejian Ren and Zixing Zhang and Qianqian Wang and Kuo Cai and Yunfan Wu and Hongtao Cheng and Zexuan Cheng and Lu Ren and Huanjie Wang and Yi Su and Ruiming Tang and Kun Gai and Guorui Zhou},
      year={2025},
      eprint={2510.11639},
      archivePrefix={arXiv},
      primaryClass={cs.IR},
      url={https://arxiv.org/abs/2510.11639}, 
}

@misc{yang2025qwen3technicalreport,
      title={Qwen3 Technical Report}, 
      author={An Yang and Anfeng Li and Baosong Yang and Beichen Zhang and Binyuan Hui and Bo Zheng and Bowen Yu and Chang Gao and Chengen Huang and Chenxu Lv and Chujie Zheng and Dayiheng Liu and Fan Zhou and Fei Huang and Feng Hu and Hao Ge and Haoran Wei and Huan Lin and Jialong Tang and Jian Yang and Jianhong Tu and Jianwei Zhang and Jianxin Yang and Jiaxi Yang and Jing Zhou and Jingren Zhou and Junyang Lin and Kai Dang and Keqin Bao and Kexin Yang and Le Yu and Lianghao Deng and Mei Li and Mingfeng Xue and Mingze Li and Pei Zhang and Peng Wang and Qin Zhu and Rui Men and Ruize Gao and Shixuan Liu and Shuang Luo and Tianhao Li and Tianyi Tang and Wenbiao Yin and Xingzhang Ren and Xinyu Wang and Xinyu Zhang and Xuancheng Ren and Yang Fan and Yang Su and Yichang Zhang and Yinger Zhang and Yu Wan and Yuqiong Liu and Zekun Wang and Zeyu Cui and Zhenru Zhang and Zhipeng Zhou and Zihan Qiu},
      year={2025},
      eprint={2505.09388},
      archivePrefix={arXiv},
      primaryClass={cs.CL},
      url={https://arxiv.org/abs/2505.09388}, 
}

@misc{chen2026groundedtokeninitializationnew,
      title={Grounded Token Initialization for New Vocabulary in LMs for Generative Recommendation}, 
      author={Daiwei Chen and Zhoutong Fu and Chengming Jiang and Haichao Zhang and Ran Zhou and Tan Wang and Chunnan Yao and Guoyao Li and Rui Cai and Yihan Cao and Ruijie Jiang and Fedor Borisyuk and Jianqiang Shen and Jingwei Wu and Ramya Korlakai Vinayak},
      year={2026},
      eprint={2604.02324},
      archivePrefix={arXiv},
      primaryClass={cs.CL},
      url={https://arxiv.org/abs/2604.02324}, 
}

@inproceedings{Lin2026tcrec,
author = {Lin, Fake and Hu, Binbin and Zheng, Zhi and Zhu, Xi and Liu, Ziqi and Zhang, Zhiqiang and Zhou, Jun and Xu, Tong},
title = {Token-level Collaborative Alignment for LLM-based Generative Recommendation},
year = {2026},
isbn = {9798400723070},
publisher = {Association for Computing Machinery},
address = {New York, NY, USA},
url = {https://doi.org/10.1145/3774904.3792727},
doi = {10.1145/3774904.3792727},
booktitle = {Proceedings of the ACM Web Conference 2026},
pages = {6909–6919},
numpages = {11},
location = {United Arab Emirates},
series = {WWW '26}
}

@inproceedings{zheng2024adapting,
  title={Adapting large language models by integrating collaborative semantics for recommendation},
  author={Zheng, Bowen and Hou, Yupeng and Lu, Hongyu and Chen, Yu and Zhao, Wayne Xin and Chen, Ming and Wen, Ji-Rong},
  booktitle={2024 IEEE 40th International Conference on Data Engineering (ICDE)},
  pages={1435--1448},
  year={2024},
  organization={IEEE}
}

@article{dong2025ctp,
author = {Dong, Zhiang and Hu, Liya and Chen, Jingyuan and Wang, Zhihua and Wu, Fei},
title = {Comprehend Then Predict: Prompting Large Language Models for Recommendation with Semantic and Collaborative Data},
year = {2025},
issue_date = {September 2025},
publisher = {Association for Computing Machinery},
address = {New York, NY, USA},
volume = {43},
number = {5},
issn = {1046-8188},
url = {https://doi.org/10.1145/3716499},
doi = {10.1145/3716499},
journal = {ACM Trans. Inf. Syst.},
month = jul,
articleno = {115},
numpages = {26}
}

@inproceedings{li2024large,
  title={Large language models for generative recommendation: A survey and visionary discussions},
  author={Li, Lei and Zhang, Yongfeng and Liu, Dugang and Chen, Li},
  booktitle={Proceedings of the 2024 joint international conference on computational linguistics, language resources and evaluation (LREC-COLING 2024)},
  pages={10146--10159},
  year={2024}
}

@inproceedings{wang2024learnable,
  title={Learnable item tokenization for generative recommendation},
  author={Wang, Wenjie and Bao, Honghui and Lin, Xinyu and Zhang, Jizhi and Li, Yongqi and Feng, Fuli and Ng, See-Kiong and Chua, Tat-Seng},
  booktitle={Proceedings of the 33rd ACM International Conference on Information and Knowledge Management},
  pages={2400--2409},
  year={2024}
}

@inproceedings{geng2022recommendation,
  title={Recommendation as language processing (rlp): A unified pretrain, personalized prompt \& predict paradigm (p5)},
  author={Geng, Shijie and Liu, Shuchang and Fu, Zuohui and Ge, Yingqiang and Zhang, Yongfeng},
  booktitle={Proceedings of the 16th ACM conference on recommender systems},
  pages={299--315},
  year={2022}
}

@inproceedings{hou2023learning,
  title={Learning vector-quantized item representation for transferable sequential recommenders},
  author={Hou, Yupeng and He, Zhankui and McAuley, Julian and Zhao, Wayne Xin},
  booktitle={Proceedings of the ACM Web Conference 2023},
  pages={1162--1171},
  year={2023}
}

@inproceedings{singh2024better,
  title={Better generalization with semantic ids: A case study in ranking for recommendations},
  author={Singh, Anima and Vu, Trung and Mehta, Nikhil and Keshavan, Raghunandan and Sathiamoorthy, Maheswaran and Zheng, Yilin and Hong, Lichan and Heldt, Lukasz and Wei, Li and Tandon, Devansh and others},
  booktitle={Proceedings of the 18th ACM Conference on Recommender Systems},
  pages={1039--1044},
  year={2024}
}

@inproceedings{ju2025generative,
  title={Generative Recommendation with Semantic IDs: A Practitioner's Handbook},
  author={Ju, Clark Mingxuan and Collins, Liam and Neves, Leonardo and Kumar, Bhuvesh and Wang, Louis Yufeng and Zhao, Tong and Shah, Neil},
  booktitle={Proceedings of the 34th ACM International Conference on Information and Knowledge Management},
  pages={6420--6425},
  year={2025}
}

@misc{hou2026surveygenerativerecommendationdata,
      title={A Survey on Generative Recommendation: Data, Model, and Tasks}, 
      author={Min Hou and Le Wu and Yuxin Liao and Yonghui Yang and Zhen Zhang and Yu Wang and Changlong Zheng and Han Wu and Richang Hong},
      year={2026},
      eprint={2510.27157},
      archivePrefix={arXiv},
      primaryClass={cs.IR},
      url={https://arxiv.org/abs/2510.27157}, 
}

@article{qu2025tokenrec,
  title={Tokenrec: Learning to tokenize id for llm-based generative recommendations},
  author={Qu, Haohao and Fan, Wenqi and Zhao, Zihuai and Li, Qing},
  journal={IEEE Transactions on Knowledge and Data Engineering},
  year={2025},
  publisher={IEEE}
}

@article{he2026reasoning,
  title={Reasoning over semantic ids enhances generative recommendation},
  author={He, Yingzhi and Sun, Yan and Tan, Junfei and Chen, Yuxin and Kong, Xiaoyu and Shen, Chunxu and Wang, Xiang and Zhang, An and Chua, Tat-Seng},
  journal={arXiv preprint arXiv:2603.23183},
  year={2026}
}

@inproceedings{penha2025semantic,
  title={Semantic ids for joint generative search and recommendation},
  author={Penha, Gustavo and D'Amico, Edoardo and De Nadai, Marco and Palumbo, Enrico and Tamborrino, Alexandre and Vardasbi, Ali and Lefarov, Max and Lin, Shawn and Heath, Timothy and Fabbri, Francesco and others},
  booktitle={Proceedings of the Nineteenth ACM Conference on Recommender Systems},
  pages={1296--1301},
  year={2025}
}

@inproceedings{he2026plum,
author = {He, Ruining and Heldt, Lukasz and Hong, Lichan and Keshavan, Raghunandan and Mao, Shifan and Mehta, Nikhil and Su, Zhengyang and Tsai, Alicia and Wang, Yueqi and Wang, Shao-Chuan and Yi, Xinyang and Baugher, Lexi and Cakici, Baykal and Chi, Ed and Goodrow, Cristos and Han, Ningren and Ma, He and Rosales, Romer and Soest, Abby Van and Tandon, Devansh and Wu, Su-Lin and Yang, Weilong and Zheng, Yilin},
title = {PLUM: Adapting Pre-trained Language Models for Industrial-scale Generative Recommendations},
year = {2026},
isbn = {9798400723070},
publisher = {Association for Computing Machinery},
address = {New York, NY, USA},
url = {https://doi.org/10.1145/3774904.3792802},
doi = {10.1145/3774904.3792802},
booktitle = {Proceedings of the ACM Web Conference 2026},
pages = {8093–8104},
numpages = {12},
location = {United Arab Emirates},
series = {WWW '26}
}

@misc{he2026implicitreasoninglargelanguage,
      title={Implicit Reasoning for Large Language Model-based Generative Recommendation}, 
      author={Yinhan He and Liam Collins and Bhuvesh Kumar and Jundong Li and Neil Shah and Donald Loveland},
      year={2026},
      eprint={2606.14142},
      archivePrefix={arXiv},
      primaryClass={cs.CL},
      url={https://arxiv.org/abs/2606.14142}, 
}

@inproceedings{Yu_2026,
   title={ThinkRec: Thinking-based recommendation via LLM},
   url={http://dx.doi.org/10.1145/3774904.3792070},
   DOI={10.1145/3774904.3792070},
   booktitle={Proceedings of the ACM Web Conference 2026},
   publisher={ACM},
   author={Yu, Qihang and Fu, Kairui and Lv, Zheqi and Zhang, Shengyu and Wu, Xinhui and Lin, Chen and Wei, Feng and Zheng, Bo and Wu, Fei},
   year={2026},
   month=Apr, pages={5698–5709} }

@inproceedings{Dobler_2023,
   title={FOCUS: Effective Embedding Initialization for Monolingual Specialization of Multilingual Models},
   url={http://dx.doi.org/10.18653/v1/2023.emnlp-main.829},
   DOI={10.18653/v1/2023.emnlp-main.829},
   booktitle={Proceedings of the 2023 Conference on Empirical Methods in Natural Language Processing},
   publisher={Association for Computational Linguistics},
   author={Dobler, Konstantin and de Melo, Gerard},
   year={2023},
   pages={13440–13454} }

@inproceedings{loveland2025role,
  title={On the Role of Weight Decay in Collaborative Filtering: A Popularity Perspective},
  author={Loveland, Donald and Ju, Mingxuan and Zhao, Tong and Shah, Neil and Koutra, Danai},
  booktitle={Proceedings of the 31st ACM SIGKDD Conference on Knowledge Discovery and Data Mining V. 2},
  pages={1975--1986},
  year={2025}
}

@inproceedings{mansoury2020recsys,
author = {Mansoury, Masoud and Abdollahpouri, Himan and Pechenizkiy, Mykola and Mobasher, Bamshad and Burke, Robin},
title = {Feedback Loop and Bias Amplification in Recommender Systems},
year = {2020},
isbn = {9781450368599},
publisher = {Association for Computing Machinery},
address = {New York, NY, USA},
url = {https://doi.org/10.1145/3340531.3412152},
doi = {10.1145/3340531.3412152},
booktitle = {Proceedings of the 29th ACM International Conference on Information \& Knowledge Management},
pages = {2145–2148},
numpages = {4},
location = {Virtual Event, Ireland},
series = {CIKM '20}
}

@inproceedings{lin2025resn,
author = {Lin, Siyi and Gao, Chongming and Chen, Jiawei and Zhou, Sheng and Hu, Binbin and Feng, Yan and Chen, Chun and Wang, Can},
title = {How Do Recommendation Models Amplify Popularity Bias? An Analysis from the Spectral Perspective},
year = {2025},
isbn = {9798400713293},
publisher = {Association for Computing Machinery},
address = {New York, NY, USA},
url = {https://doi.org/10.1145/3701551.3703579},
doi = {10.1145/3701551.3703579},
booktitle = {Proceedings of the Eighteenth ACM International Conference on Web Search and Data Mining},
pages = {659–668},
numpages = {10},
location = {Hannover, Germany},
series = {WSDM '25}
}

@inproceedings{liu2023text,
  title={Text matching improves sequential recommendation by reducing popularity biases},
  author={Liu, Zhenghao and Mei, Sen and Xiong, Chenyan and Li, Xiaohua and Yu, Shi and Liu, Zhiyuan and Gu, Yu and Yu, Ge},
  booktitle={Proceedings of the 32nd ACM international conference on information and knowledge management},
  pages={1534--1544},
  year={2023}
}

@inproceedings{xi2024towards,
  title={Towards open-world recommendation with knowledge augmentation from large language models},
  author={Xi, Yunjia and Liu, Weiwen and Lin, Jianghao and Cai, Xiaoling and Zhu, Hong and Zhu, Jieming and Chen, Bo and Tang, Ruiming and Zhang, Weinan and Yu, Yong},
  booktitle={Proceedings of the 18th ACM Conference on Recommender Systems},
  pages={12--22},
  year={2024}
}

@article{kusupati2022matryoshka,
  title={Matryoshka representation learning},
  author={Kusupati, Aditya and Bhatt, Gantavya and Rege, Aniket and Wallingford, Matthew and Sinha, Aditya and Ramanujan, Vivek and Howard-Snyder, William and Chen, Kaifeng and Kakade, Sham and Jain, Prateek and others},
  journal={Advances in Neural Information Processing Systems},
  volume={35},
  pages={30233--30249},
  year={2022}
}

@article{nussbaumnomic,
  title={Nomic Embed: Training a Reproducible Long Context Text Embedder},
  author={Nussbaum, Zach and Morris, John Xavier and Mulyar, Andriy and Duderstadt, Brandon},
  journal={Transactions on Machine Learning Research},
  year={2024}
}

@article{yamaguchi2026,
    title = "How Can We Effectively Expand the Vocabulary of {LLM}s with 0.01{GB} of Target Language Text?",
    author = "Yamaguchi, Atsuki  and
      Villavicencio, Aline  and
      Aletras, Nikolaos",
    journal = "Computational Linguistics",
    volume = "52",
    number = "1",
    month = mar,
    year = "2026",
    address = "Cambridge, MA",
    publisher = "MIT Press",
    url = "https://aclanthology.org/2026.cl-1.9/",
    doi = "10.1162/coli.a.581",
    pages = "295--330",
}

@inproceedings{mundra2024empirical,
    title = "An Empirical Comparison of Vocabulary Expansion and Initialization Approaches For Language Models",
    author = "Mundra, Nandini  and
      Khandavally, Aditya Nanda Kishore  and
      Dabre, Raj  and
      Puduppully, Ratish  and
      Kunchukuttan, Anoop  and
      Khapra, Mitesh M",
    editor = "Barak, Libby  and
      Alikhani, Malihe",
    booktitle = "Proceedings of the 28th Conference on Computational Natural Language Learning",
    month = nov,
    year = "2024",
    address = "Miami, FL, USA",
    publisher = "Association for Computational Linguistics",
    url = "https://aclanthology.org/2024.conll-1.8/",
    doi = "10.18653/v1/2024.conll-1.8",
    pages = "84--104",
}

@article{chen2026llms,
  title={LLMs Need Encoders for Semantic IDs Too},
  author={Chen, Xiangyi and Wang, Zelun and Li, Xinyi and Hsu, Yi-Ping and Yang, Jaewon and Xu, Jiajing},
  journal={arXiv preprint arXiv:2606.00324},
  year={2026}
}

@article{li2026qwen3,
  title={Qwen3-VL-Embedding and Qwen3-VL-Reranker: A Unified Framework for State-of-the-Art Multimodal Retrieval and Ranking},
  author={Li, Mingxin and Zhang, Yanzhao and Long, Dingkun and Chen, Keqin and Song, Sibo and Bai, Shuai and Yang, Zhibo and Xie, Pengjun and Yang, An and Liu, Dayiheng and others},
  journal={arXiv preprint arXiv:2601.04720},
  year={2026}
}

@article{zhou2025openonerec,
  title={OpenOneRec Technical Report},
  author={Zhou, Guorui and Bao, Honghui and Huang, Jiaming and Deng, Jiaxin and Zhang, Jinghao and She, Junda and Cai, Kuo and Ren, Lejian and Ren, Lu and Luo, Qiang and others},
  journal={arXiv preprint arXiv:2512.24762},
  year={2025}
}

@article{
yang2025unifying,
title={Unifying Generative and Dense Retrieval for Sequential Recommendation},
author={Liu Yang and Fabian Paischer and Kaveh Hassani and Jiacheng Li and Shuai Shao and Zhang Gabriel Li and Yun He and Xue Feng and Nima Noorshams and Sem Park and Bo Long and Robert D Nowak and Xiaoli Gao and Hamid Eghbalzadeh},
journal={Transactions on Machine Learning Research},
issn={2835-8856},
year={2025},
note={}
}

\appendix

\section{Additional Related Work}
\label{sec:related}

In this section, we expand on the related work introduced throughout the main text. We focus on two core sections: generative recommendation paradigms and alignment strategies, which most closely resemble our work.

\vspace{0.1cm}
\noindent \textbf{Generative Recommendation Paradigms.}
While this work focuses specifically on LLM-based GR pipelines, the broader generative recommendation landscape encompasses a wide array of architectures. In their simplest form, from-scratch sequential recommenders, such as HSTU, recast recommendation as a sequential transduction over raw action sequences \cite{zhai2024actions}. To accommodate larger catalogs, TIGER established the use of SIDs to represent items, training an encoder-decoder transformer from scratch \cite{rajput2023recommender}. Building on this, EAGER introduces parallel behavior and semantic decoders \cite{wang2024eager}, while the OneRec family scales to industrial settings using unified encoder-decoder architectures refined through iterative preference alignment \cite{deng2025onerecunifyingretrieverank, MiniOneRec, zhou2025openonerec}. Closest to our setting, PLUM adapts pretrained LLMs for YouTube-scale recommendation using the exact multi-stage recipe (SID tokenization, CPT, and task fine-tuning) that our study dissects \cite{he2026plum}. Orthogonally, a parallel line of work leverages LLMs without generating item identifiers, e.g., acting as zero-shot rankers \cite{hou2024large}, prompt-tuned predictors \cite{dong2025ctp}, or external knowledge generators \cite{xi2024towards}. Throughout the evolution of these designs, a primary motivation for adopting pretrained LLMs over from-scratch architectures is their capacity for rich semantic understanding. While our evaluation centers on standard LLM-based GR pipelines to better evoke their semantic capabilities, our findings highlight a more fundamental principle: whenever SIDs are integrated into a GR model, preserving their semantic geometry at initialization can be a valuable means of fully realizing this semantic promise, rather than allowing the architecture to regress to standard collaborative biases.

\vspace{0.1cm}
\noindent \textbf{Aligning SID Tokens with the LLM Vocabulary.}
Given the prevailing strategy of randomly initializing SID token embeddings, prior works have proposed increasingly complex training pipelines to align this new vocabulary with the LLM. For instance, LC-Rec introduced auxiliary alignment tasks to ground SIDs in language and collaborative semantics \cite{zheng2024adapting}, a recipe that more recent pipelines have consolidated into their CPT (itemic alignment) stage \cite{he2026plum}. Closest to our diagnostic analysis, GTI similarly observes that mean-initialized SID tokens collapse into a degenerate subspace \cite{chen2026groundedtokeninitializationnew}. However, while GTI identifies this structural collapse (measured via effective rank), its proposed remedy is an entirely separate grounding stage that trains the new token rows over paired item-text prompts. This exact strategy is also adopted via OneRec-Think in their CPT stage \cite{liu2025onerecthinkintextreasoninggenerative}. In contrast, our proposed centroid initialization directly reuses the continuous geometry already established during SID construction, providing a one-time operation that bypasses the need for auxiliary gradient steps, specialized data curation, or additional tuning. 

Beyond explicit pre-training, other alignment mechanisms introduce large architectural or decoding complexities to the GR pipeline. TCA4Rec injects collaborative signals through soft next-token labels distilled from a teacher model \cite{Lin2026tcrec}, while Prefixmem treats SIDs as a distinct modality by attaching a prefix-conditioned encoder \cite{chen2026llms}. More recently, a wave of research has focused on eliciting reasoning around SID generation, whether through explicit reasoning traces \cite{Yu_2026}, reinforcement learning \cite{he2026reasoning}, or implicit reasoning \cite{he2026implicitreasoninglargelanguage}. Finally, hybrid architectures such as LIGER interpolate between dense and generative retrieval, in part to recover the cold-start performance that pure SID decoding typically loses \cite{yang2025unifying}. However, our findings trace this cold-start gap to the semantic geometry discarded during standard vocabulary expansion. Thus, by resolving this directly at initialization, our approach improves tail-item performance without relying on auxiliary encoders, teacher models, or hybrid dense towers.

\section{Datasets, Hyperparameters, and \\Experimental Details}
\label{sec:hyperparams}

\begin{table}[t]\centering
\caption{Statistics for datasets used throughout the paper.}
\label{tab:stats}
\begin{tabular}{lcccc}
\toprule
Domain & users & items & interactions & avg.\ len \\
\midrule
Beauty & 22363 & 12101 & 194687 & 8.7 \\
Sports & 35598 & 18357 & 249017 & 7.0 \\
Toys & 19412 & 11924 & 136091 & 7.0 \\
Steam  & 47761 & 12012 & 599620 & 12.6 \\
MovieLens & 50000 & 26744 & 7212178 & 144.2 \\
\bottomrule
\end{tabular}
\end{table}
\footnotetext{For MovieLens, we use a 50K-user subsample of ML-20M.}

In this section, we detail our experimental setup. \Cref{tab:hyper} summarizes the shared settings across all experiments, with individual runs differing only in their dataset, backbone scale, SID embedding initialization (\Cref{eq:randinit} vs.\ \Cref{eq:centinit}), and number of CPT epochs.

\begin{table}[t]\centering
\caption{\textbf{Hyperparameter and Experimental Setup.}}
\label{tab:hyper}
\small
\setlength{\tabcolsep}{4pt}
\begin{tabular}{ll}
\toprule
\multicolumn{2}{l}{\textbf{Backbone and Vocabulary Expansion}} \\
Backbone & Qwen3 (of varying size)\\
Added Tokens & $4{\times}256$ SID codes (\sid{<s\_a\_k>}--\sid{<s\_d\_k>}) \\
 & $+$ 2 delimiters (\sid{<|sid\_begin|>}, \sid{<|sid\_end|>}) \\
Random Init. & HF \texttt{resize\_token\_embeddings}, mean resizing \\
Centroid Init. & \Cref{eq:centinit} on the $3{\times}256$ semantic codes; \\
 & dedup codes and delimiters remain random \\
\midrule
\multicolumn{2}{l}{\textbf{Semantic embeddings and SIDs}} \\
Content Encoder & Qwen3-Embedding-8B \\
Dimensionality & 4096 Native, Matryoshka-truncated to $d_{\mathrm{model}}$ \\
Quantizer & Residual $k$-means, $L{=}3$ semantic levels, $K{=}256$ \\
\midrule
\multicolumn{2}{l}{\textbf{CPT (Phase 1, Alignment)}} \\
Epochs / LR / decay & $\le$10 / $10^{-4}$ (AdamW) / 0 \\
Effective Batch & 64 \\
Precision & BF16 \\
Trainable Params & Newly added rows only (PEFT TrainableTokens) \\
LR Warmup & None \\
Max Seq.\ Length & 4096 \\
Grad.\ Clipping & Max Norm 1.0  \\
\midrule
\multicolumn{2}{l}{\textbf{SFT (Phase 2, Next-item Prediction)}} \\
Epochs / LR / decay & $\le$12 / $10^{-5}$ (AdamW) / 0.01 \\
Effective batch & 16  \\
Precision & FP32 Weights, BF16-mixed for Optimizer \\
Trainable Params & Full Fine-Tune \\
LR Warmup & Starting LR 10\%, Then Constant \\
Max Seq.\ Length & 1024 \\
Grad.\ Clipping & Max Norm 1.0  \\
\midrule
\multicolumn{2}{l}{\textbf{Evaluation}} \\
Decoding & Trie-constrained Beam Search \\
Beams & 10 \\
\bottomrule
\end{tabular}
\end{table}

\vspace{0.1cm}
\noindent \textbf{Datasets and Preprocessing.}
We evaluate our approach across five distinct datasets (\Cref{tab:stats}). The three Amazon product datasets (Beauty, Sports, Toys) follow the standard 5-core preprocessing used in prior GR work \cite{geng2022recommendation, rajput2023recommender}, where each item is described by its title, categorization, and price. Steam contains video-game interactions in which each game carries a title and a list of community tags. For MovieLens, we subsample 50{,}000 users from ML-20M. For scalability, we truncate all item sequences to their most recent 21 interactions. All datasets use the standard leave-one-out protocol, where, for each user, the final interaction is the test target, the penultimate interaction is the validation target, and all preceding interactions form the training sequence.

\vspace{0.1cm}
\noindent \textbf{Semantic Embeddings.}
Item semantic embeddings are generated by Qwen3-Embedding-8B \cite{yang2025qwen3technicalreport}. Each item's text (\Cref{sec:prompts}) is passed into the encoder with no task instruction, tokenized to a maximum length of 256 with right padding, and embedded by extracting the final-layer hidden state at the last non-padding token. Exploiting the encoder's Matryoshka property \cite{kusupati2022matryoshka}, the 4096-dimensional output is truncated to its leading 2048 dimensions prior to quantization, ensuring the centroids directly match the Qwen3-1.7B embedding width. For the Qwen3-0.6B scale study, these identical centroids are further truncated to their leading 1024 dimensions. 

\vspace{0.1cm}
\noindent \textbf{SID Construction.}
SIDs are generated via residual $k$-means over the truncated item embeddings using $L{=}3$ semantic levels of $K{=}256$ codes, with residual normalization at each level (\Cref{eq:rq}). For Beauty, Sports, and Toys, each level is fit using mini-batch $k$-means ($k$-means++ initialization, 3{,}000 optimization steps at a batch size of 2{,}048). For Steam and MovieLens, each level is fit with full-batch Lloyd's algorithm run to convergence, yielding better code utilization. Collisions among the 3-code semantic prefixes are resolved by the deterministic deduplication code $k_L(i)$, resulting in a strictly unique 4-code SID per item. 

\vspace{0.1cm}
\noindent \textbf{Vocabulary Expansion and Initialization.}
The tokenizer is expanded with 1{,}026 special tokens: \sid{<s\_a\_k>}, \sid{<s\_b\_k>}, and \sid{<s\_c\_k>} represent the semantic levels $\ell=0,1,2$, respectively, while \sid{<s\_d\_k>} represents the deduplication code $k_L(i)$; each uses $k\in\{0,\dots,K-1\}$ with $K=256$. We additionally introduce the delimiter tokens \sid{<|sid\_begin|>} and \sid{<|sid\_end|>}. The random, mean-initialized baseline uses the standard HuggingFace mean-resizing default, initializing each new row from the scaled Gaussian defined in \Cref{eq:randinit}. Centroid initialization overwrites the semantic levels $\ell\in\{0,\dots,L-1\}$ via \Cref{eq:centinit}, computing $\boldsymbol{\mu}_{\mathrm{base}}$ over the original vocabulary rows $\mathcal{V}_{\mathrm{base}}$. Deduplication codes and delimiters retain their mean-resized initialization. As the Qwen3 architecture ties its input embeddings to the language-modeling head, this write initializes both the input and output representations for every SID token.

\vspace{0.1cm}
\noindent \textbf{CPT Configuration.}
The CPT phase trains for up to 10 epochs using AdamW ($\beta = (0.9, 0.999)$, $\epsilon = 10^{-8}$) at a constant learning rate of $10^{-4}$, employing no weight decay and no warmup. Only the 1{,}026 newly added embedding rows are trainable (via PEFT TrainableTokens). The effective batch size is 64. Sequences are capped at 4{,}096 tokens, and gradient checkpointing is enabled. Training examples consist of item-interaction histories rendered as interleaved SID-text descriptions (\Cref{sec:prompts}) optimized with a standard next-token objective, training the LLM to generate an item’s text from its accompanying SID. Note that this prompting strategy and objective directly match OneRec-Think's CPT \cite{liu2025onerecthinkintextreasoninggenerative} and the user behavior CPT for PLUM \cite{he2026plum}. For the CPT-epoch sweeps, SFT is performed from intermediate checkpoints after $\{1,4,6,8,10\}$ CPT epochs. 

\vspace{0.1cm}
\noindent \textbf{SFT Configuration.}
The SFT phase trains all model parameters for up to 12 epochs (with early stopping monitored via validation Recall@5, patience 5) using AdamW at a learning rate of $10^{-5}$ and a weight decay of 0.01. The learning rate schedule uses a warmup ratio of 0.1 followed by a constant rate. Master weights are stored in fp32 with bf16-mixed autocast, and gradients are clipped to a max-norm of 1.0. The effective batch size is 16. Prompts adhere to the ChatML format detailed in \Cref{sec:prompts}, again following OneRec-Think \cite{liu2025onerecthinkintextreasoninggenerative}. To focus on generation, the loss is computed only from the user turn onward, masking the system prompt. Sequences are capped at 1{,}024, utilizing left-truncation to preserve the most recent interactions.

\vspace{0.1cm}
\noindent \textbf{Evaluation Protocol.}
Item generation is executed via constrained beam search over a trie built on each dataset's valid item SIDs. At each generation step, the logits of tokens that do not extend a valid SID prefix are masked to $-\infty$, guaranteeing that every generated sequence corresponds to a real item. For performance, we report Recall@$k$, which measures whether the ground-truth SID is present among the top-$k$ beams. Following standard practice, we monitor Recall after every SFT epoch using 10 beams evaluated on the leave-one-out validation targets and retain the checkpoint with the highest validation Recall@5. All final evaluations are performed on the test targets using the same 10-beam configuration. The convergence step $t^\star$ reported in \Cref{tab:av1} is the exact optimizer step at which this validation monitor peaks.

\vspace{0.1cm}
\noindent \textbf{Hardware.}
All training and evaluation are performed with 4 NVIDIA A100-80GB GPUs. Under this  setup, a full CPT phase can take up to approximately one day per dataset, highlighting the practical value of strategies that can reduce or eliminate the need for CPT.

\subsection{Latent Signal Encoding Targets}
\label{sec:latent_signal_targets}

In this section, we formalize the construction of the target matrices $\mathbf{T}_{\mathrm{pop}}$ and $\mathbf{T}_{\mathrm{sem}}$ used for the latent signal encoding analysis presented in \Cref{sec:prop_enc}. Both targets are defined over the $K{=}256$ level-0 SID codes associated with $\mathcal{C}_0$.

\vspace{0.1cm}
\noindent \textbf{Popularity Target.}
Let $f_i$ denote the number of training examples whose ground-truth next item is $i$. Then, the aggregate popularity of level-0 code $k$ sums this frequency over all items assigned to it. To stabilize variance against the heavy-tailed frequency distribution, the popularity target applies a logarithmic transform:
\begin{equation}
n_k = \sum_{i \,:\, k_0(i) = k} f_i,
\qquad
\mathbf{T}_{\mathrm{pop}} = \big[\log(1 + n_k)\big]_{k=0}^{K-1} \in \mathbb{R}^{K \times 1}.
\label{eq:tpop}
\end{equation}

\vspace{0.1cm}
\noindent \textbf{Semantic Target.}
Assuming each item carries a single categorical label $\mathrm{cat}(i)$ (detailed below), a level-0 code inherits the majority label of its assigned items:
\begin{equation}
\mathrm{cat}_0(k) = \operatorname*{arg\,max}_{c} \; \big|\{\, i : k_0(i) = k,\; \mathrm{cat}(i) = c \,\}\big|.
\label{eq:tsem-dom}
\end{equation}
The semantic target is then defined as the one-hot indicator matrix over the $C_{\mathrm{cat}}$ observed code-level categories, yielding $\mathbf{T}_{\mathrm{sem}} \in \mathbb{R}^{K \times C_{\mathrm{cat}}}$ where $[\mathbf{T}_{\mathrm{sem}}]_{k,c} = \mathbb{1}[\mathrm{cat}_0(k) = c]$. Codes devoid of assigned items contribute zero vectors. 

To ensure $R_q^2(\mathbf{T})$ strictly measures variance captured around the mean, both targets are column-centered ($\mathbf{T} \leftarrow \mathbf{T} - \bar{\mathbf{T}}$) prior to evaluation. The basis $\mathbf{Z}_q$ is derived from the identically centered embedding matrix. These same code-level labels $\mathrm{cat}_0(k)$ are also used to compute the neighborhood purity metric $P_m$.

\vspace{0.1cm}
\noindent \textbf{Category Construction per Dataset.}
For the Amazon domains, $\mathrm{cat}(i)$ is the provided coarse category (e.g., extracting \emph{Hair Care} from ``\emph{Hair Care} $>$ Conditioners'').  For MovieLens, $\mathrm{cat}(i)$ maps to the first genre in the item's genre list. Steam, by contrast, lacks a formal categorical taxonomy and instead carries free-form community tag lists. To derive a taxonomy, we segment each tag string by matching against a curated lexicon of known Steam tags. Once the tag strings are found, we discard any descriptive modifier tags (e.g., \emph{Indie}, \emph{Free to Play}, \emph{Singleplayer}, \emph{2D}). Then, exploiting Steam's relevance-ordered tag convention, the first remaining tag is mapped into one of 11 genre buckets: Shooter, RPG, Strategy, Simulation, Action, Adventure, Puzzle \& Casual, Horror \& Survival, Sports \& Racing, MMO, and Software. Items without any matched content tag are labeled {Other} and excluded from purity computations.

\definecolor{sidpurple}{HTML}{5315F5}   
\definecolor{itemgreen}{HTML}{1A7A4A}   
\definecolor{ctrlorange}{HTML}{B34700}  
\definecolor{promptbg}{RGB}{248,248,248}
\newcommand{\ptok}[1]{\textcolor{sidpurple}{\texttt{#1}}}
\newcommand{\pitem}[1]{\textcolor{itemgreen}{\texttt{#1}}}
\newcommand{\pctrl}[1]{\textcolor{ctrlorange}{\texttt{#1}}}
\newcommand{\promptbox}[1]{%
  \noindent\fcolorbox{black!25}{promptbg}{%
    \parbox{0.94\columnwidth}{\raggedright\footnotesize\ttfamily #1}}}

\section{Example Prompts}
\label{sec:prompts}

In this section, we detail the prompt templates and provide examples with real items from Amazon Beauty. Throughout, \pctrl{orange} denotes chat-template special tokens, \ptok{purple} denotes SID tokens, \pitem{green} denotes item text.

\vspace{0.1cm}
\noindent \textbf{Semantic Embedding Input.}
The input to Qwen3-Embedding-8B is the item's text, passed with no task instruction (\Cref{fig:prompt-embed}). Item embeddings are read from the final-layer hidden state at the last non-padding token.

\begin{figure}[t]
\centering
\begin{tcolorbox}[colback=gray!5!white,colframe=gray!75!black,title=Semantic Embedding Input,halign=flush left]
\small
Title: \pitem{OPI Nail Lacquer, Simmer and Shimmer, 0.5-Fluid Ounce}; Brand: \pitem{OPI}; Categories: [\pitem{'Beauty', 'Makeup', 'Nails', 'Nail Polish'}]; Price: \pitem{12.0};
\end{tcolorbox}
\vspace{-0.3cm}
\caption{Input to the embedding model. For Beauty, this includes an item's title, brand, category path, and price, without any instruction.}
\label{fig:prompt-embed}
\end{figure}

\vspace{0.1cm}
\noindent \textbf{CPT Prompt.}
CPT examples are plain text sequences that interleave each item's SID with its title and categories over item-interaction histories (\Cref{fig:prompt-cpt}), trained with the standard next-token objective. This matches the CPT objectives in OneRec-Think \cite{liu2025onerecthinkintextreasoninggenerative} and PLUM \cite{he2026plum}.

\begin{figure}[t]
\centering
\begin{tcolorbox}[colback=gray!5!white,colframe=gray!75!black,title=CPT Prompt Structure,halign=flush left]
\small
The user has purchased the following items: \ptok{<|sid\_begin|>\allowbreak<s\_a\_147>\allowbreak<s\_b\_129>\allowbreak<s\_c\_3>\allowbreak<s\_d\_0>\allowbreak<|sid\_end|>}, its title is "\pitem{Phyto Phytocitrus Restructuring Mask for Unisex, 6.7 Ounce}", its categories are "\pitem{Beauty > Hair Care > Conditioners}"; \ptok{<|sid\_begin|>\allowbreak<s\_a\_45>\allowbreak<s\_b\_12>\allowbreak<s\_c\_231>\allowbreak<s\_d\_0>\allowbreak<|sid\_end|>}, its title is "\pitem{Matrix Biolage Colorcaretherapie Color Care Shampoo and Conditioner Set 33.8oz 1 Liter}", its categories are "\pitem{Beauty > Hair Care > Shampoo \& Conditioner Sets}"; \ptok{<|sid\_begin|>\allowbreak<s\_a\_130>\allowbreak<s\_b\_111>\allowbreak<s\_c\_88>\allowbreak<s\_d\_3>\allowbreak<|sid\_end|>}, its title is "\pitem{Raw African Black Soap from Ghana 1 Lb}", its categories are "\pitem{Beauty > Bath \& Body > Cleansers > Soaps}"; \dots;
\end{tcolorbox}
\vspace{-0.3cm}
\caption{CPT example for Amazon Beauty: A plain causal-LM sequence interleaving each history item's SID with its title and category path.}
\label{fig:prompt-cpt}
\end{figure}

\vspace{0.1cm}
\noindent \textbf{SFT Prompt.}
SFT examples use the ChatML format (\Cref{fig:prompt-sft}), following OneRec-Think \cite{liu2025onerecthinkintextreasoninggenerative}. The user turn contains the interaction history as a SID-only sequence, and the assistant turn contains an empty think block followed by the target item's SID. The loss is computed from the user turn onward. At inference, the prompt ends immediately after the think block, and constrained beam search generates the recommended SIDs.

\begin{figure}[t]
\centering
\begin{tcolorbox}[colback=gray!5!white,colframe=gray!75!black,title=SFT Prompt Structure,halign=flush left]
\small
\pctrl{<|im\_start|>}\pctrl{system}\\
You are a professional recommendation expert who needs to recommend the next possible purchase for users based on their purchase history. Please predict the most likely next product that the user will purchase based on the user's historical purchase information.\pctrl{<|im\_end|>}\\
\pctrl{<|im\_start|>}\pctrl{user}\\
The user has purchased the following items: \ptok{<|sid\_begin|>\allowbreak<s\_a\_147>\allowbreak<s\_b\_129>\allowbreak<s\_c\_3>\allowbreak<s\_d\_0>\allowbreak<|sid\_end|>}; \ptok{<|sid\_begin|>\allowbreak<s\_a\_45>\allowbreak<s\_b\_12>\allowbreak<s\_c\_231>\allowbreak<s\_d\_0>\allowbreak<|sid\_end|>}; \ptok{<|sid\_begin|>\allowbreak<s\_a\_130>\allowbreak<s\_b\_111>\allowbreak<s\_c\_88>\allowbreak<s\_d\_3>\allowbreak<|sid\_end|>}; \dots;\pctrl{<|im\_end|>}\\
\pctrl{<|im\_start|>}\pctrl{assistant}\\
\pctrl{<think>}\\
\pctrl{</think>}\\
\ptok{<|sid\_begin|>\allowbreak<s\_a\_46>\allowbreak<s\_b\_84>\allowbreak<s\_c\_144>\allowbreak<s\_d\_0>\allowbreak<|sid\_end|>}\pctrl{<|im\_end|>}
\end{tcolorbox}
\vspace{-0.3cm}
\caption{SFT example for Amazon Beauty: Uses ChatML format where the user turn carries the SID-only interaction history, and the assistant turn carries an empty think block followed by the ground-truth next item's SID.}
\label{fig:prompt-sft}
\end{figure}

\section{Resolving Item-Level vs.\ Token-Level \\ Popularity}
\label{sec:item-vs-token-pop}

\begin{figure*}[t]
\centering
\includegraphics[width=0.8\textwidth]{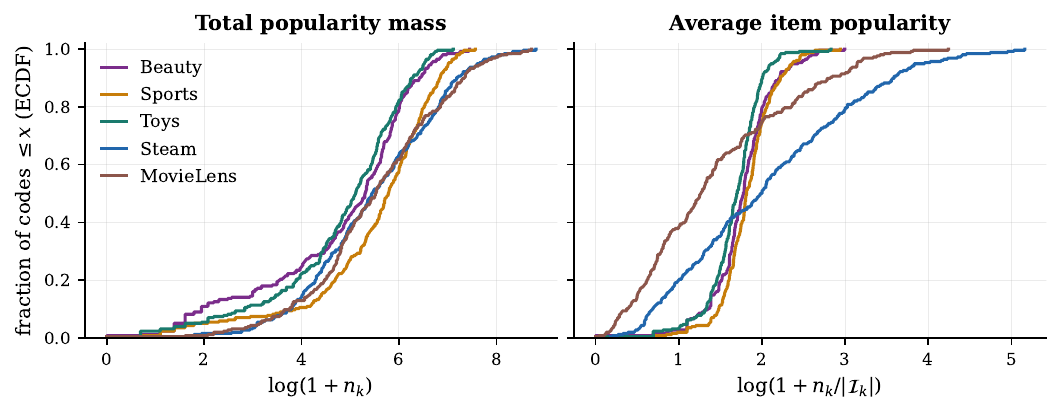}
\caption{\textbf{Token frequency vs.\ average item popularity for the level-0 codebook $\mathcal{C}_0$.} \emph{Left:} Total code popularity mass ($\log(1+n_k)$) is broadly dispersed across all datasets, confirming it as an informative geometric target. \emph{Right:} Average item popularity ($\log(1+n_k/|\mathcal{I}_k|)$) spreads widely on Steam and MovieLens. While the Amazon domains exhibit tighter concentration (steep curves), the distribution is not perfectly uniform; this subtle variance ensures that a measurable, albeit weak, popularity signal remains encoded even at level 0.}
\label{fig:pop-target-dist}
\end{figure*}

In this section, we directly address the relationship between our token-level diagnostics and item-level behavioral popularity bias. Specifically, our analyses evaluate popularity through two granularities: \emph{item-level} popularity $f_i$ and \emph{token-level} popularity $n_k = \sum_{i\in\mathcal{I}_k} f_i$, where $\mathcal{I}_k=\{i:k_0(i)=k\}$ is the set of items assigned to level-0 code $k$. 
While token-level popularity is derived from item popularity, we note that relying on a single codebook level introduces ambiguity because aggregate frequency conceals the distribution of constituent items within a code. Mathematically, level-0 code frequency factors into cluster size and average item popularity ($n_k = |\mathcal{I}_k| \cdot \bar{f}_k$). \Cref{fig:pop-target-dist} illustrates how these underlying factors behave across datasets. For instance, on Steam and MovieLens, average item popularity spreads widely, meaning a level-0 code's frequency strongly reflects the popularity of its assigned items. Conversely, on the Amazon domains, average item popularity is more concentrated, meaning variance in level-0 code frequency is a mixture of cluster size and actual item popularity. 

On the surface, this divergence might suggest that a model encoding codebook-level aggregate popularity cannot exploit item popularity, potentially severing the link between our geometric diagnostic (\Cref{sec:prop_enc}) and downstream popularity bias (\Cref{sec:tail-cold}). However, this discrepancy arises only when evaluating a single codebook level in isolation. Instead, given the LLM traverses the complete SID hierarchy during generation, it is able to implicitly learn conditional token probabilities. Then, the product of these conditional probabilities is able to recover the exact item popularity:
\begin{equation}
\frac{f_i}{N} = \prod_{j=0}^{L} p_{\mathrm{data}}\big(k_j(i) \mid k_{<j}(i)\big),
\label{eq:pop-chainrule}
\end{equation}
where $N$ is the total number of prediction examples, $p_{\mathrm{data}}$ is the empirical conditional distribution, and $k_{<j}(i)=(k_0(i),\dots,k_{j-1}(i))$ denotes the SID prefix preceding level $j$. 
To quantify this recovery, let $a_i=\log(1+f_i)$ and let $\bar a_g$ denote the mean of $a_i$ over items sharing prefix $g$. We perform a one-way variance decomposition by partitioning the catalog according to $k_{<\rho}(i)$ for each semantic-prefix length $\rho\in\{1,\dots,L\}$:
\begin{equation}
\mathrm{EV}_\rho
=
\frac{\operatorname{Var}_i\!\left[\bar a_{k_{<\rho}(i)}\right]}
{\operatorname{Var}_i[a_i]}.
\end{equation}
Intuitively, the explained-variance ratio $\mathrm{EV}_\rho$ measures the percentage of an item's log-popularity that can be explained by knowing its SID prefix. Consequently, grouping items by increasingly longer prefixes captures a steadily growing share of this variance, achieving near-recovery across the three semantic levels (e.g., $\mathrm{EV}_\rho$ on Beauty scales $0.06 \to 0.51 \to 0.86$ for prefixes through levels 0, 1, and 2, while Steam scales $0.28 \to 0.85 \to 0.99$).
This property highlights that token-level popularity and item-level popularity are equivalent signals viewed at different granularities of the generative sequence. Thus, while our codebook-level diagnostics probe just a single level, item-level popularity bias inevitably emerges as a cumulative consequence of decoding the full SID hierarchy.

\section{CPT as Popularity Debiasing}
\label{sec:cptdebias}
\begin{table}[h]
\centering
\caption{Recall@5 for Head (top 20\%) and Tail (bottom 80\%) items across CPT depths prior to SFT under random initialization. The declining Head/Tail ratio (shaded gray) demonstrates that deeper CPT mitigates popularity bias by boosting tail performance up to peak validation depth.}
\label{tab:cptdiagnostic}
\small
\setlength{\tabcolsep}{4pt}
\begin{tabular}{llcccccc}
\toprule
Dataset & Metric & no CPT & $1$ & $4$ & $6$ & $8$ & $10$ \\
\midrule
\multirow{3}{*}{Beauty} & head@5    & 0.079 & 0.097 & 0.097 & 0.098 & 0.095 & 0.099 \\
  & tail@5    & 0.017 & 0.020 & 0.024 & 0.022 & 0.023 & 0.026 \\
  \rowcolor{gray!12} \cellcolor{white} & head/tail & 4.60  & 4.80  & 3.96  & 4.37  & 4.19  & 3.76  \\
\midrule
\multirow{3}{*}{Sports} & head@5    & 0.058 & 0.063 & -- & -- & -- & -- \\
  & tail@5    & 0.007 & 0.011 & -- & -- & -- & -- \\
  \rowcolor{gray!12} \cellcolor{white} & head/tail & 8.06  & 5.71  & -- & -- & -- & -- \\
\midrule
\multirow{3}{*}{Toys} & head@5    & 0.087 & 0.088 & 0.093 & 0.093 & 0.094 & -- \\
  & tail@5    & 0.029 & 0.033 & 0.035 & 0.038 & 0.039 & -- \\
  \rowcolor{gray!12} \cellcolor{white} & head/tail & 2.99  & 2.67  & 2.65  & 2.43  & 2.41  & -- \\
\midrule
\multirow{3}{*}{Steam} & head@5    & 0.192 & 0.200 & 0.201 & 0.199 & -- & -- \\
  & tail@5    & 0.004 & 0.006 & 0.006 & 0.007 & -- & -- \\
  \rowcolor{gray!12} \cellcolor{white} & head/tail & 51.28 & 34.83 & 34.88 & 27.51 & -- & -- \\
\midrule
\multirow{3}{*}{MovieLens} & head@5    & 0.154 & 0.168 & 0.167 & 0.166 & -- & -- \\
  & tail@5    & 0.013 & 0.019 & 0.019 & 0.020 & -- & -- \\
  \rowcolor{gray!12} \cellcolor{white} & head/tail & 12.20 & 8.94  & 8.76  & 8.50  & -- & -- \\
\bottomrule
\end{tabular}
\end{table}

In \Cref{sec:tail-cold}, we noted that additional CPT epochs tend to improve performance on colder items, even under random initialization. In \Cref{tab:cptdiagnostic}, we demonstrate this trend by decomposing Recall along the popularity axis. Specifically, we partition the catalog by training frequency into a head (top 20\% of items) and a tail (bottom 80\%), assign each test example the tier of its ground-truth target, and report Recall@5 per tier for randomly initialized models across different numbers of CPT epochs. The head-to-tail Recall ratio (shaded rows) summarizes the discrepancy between these two cohorts, where a perfectly neutral model would drive this ratio toward one.

Our analysis reveals two key trends. First, \textit{head-cohort gains are modest and saturate quickly}; on Beauty, head Recall jumps from 0.079 to 0.097 after just one CPT epoch, but barely improves (0.099) after ten. Second, \textit{tail-cohort gains scale with CPT depth}, yielding 34\% to 75\% relative improvements across datasets. Consequently, deeper CPT significantly improves the head/tail performance ratio over the baseline. Ultimately, standard random initialization struggles with data-sparse tail items, forcing models to waste CPT epochs relearning missing semantics. Centroid initialization bypasses this bottleneck by supplying semantic priors upfront, driving large pure-SFT gains on tail items (\Cref{sec:tail-cold}) and accelerating overall convergence (\Cref{fig:cptdepth}).

\section{Proof for Theorem 3.1}
\label{sec:theory_app}

In this section, we provide the proof for \Cref{thm:accumulated-bias-main} by deriving the update to a
(mean-centered) SID token embedding. The high-level goal is to isolate an explicit contribution of code popularity while retaining all context- and
model-dependent effects in a residual term, which we further characterize at the end. With this in mind, the proof is meant to provide intuition on \textit{how} popularity can become encoded within the LLM embeddings, not state that it \textit{will} become encoded. 

\vspace{0.1cm}
\noindent\textbf{Setup.}
We first consider a semantic level $\ell$ with codebook
$\mathcal{C}_{\ell}$. Then, we identify each code index
$v\in\{0,\dots,K-1\}$ with its corresponding token in the expanded
LLM vocabulary. Let $\mathcal V$ denote this expanded vocabulary. At optimizer step $t$, let
$\mathbf{o}_u^{(t)}\in\R^{d_{\mathrm{model}}}$ denote the output embedding
of any token $u\in\mathcal V$. We assume that the output embedding table is
untied from the input embedding table, so each output row affects the loss
only through the output logits. 
Next, let $\mathcal Q$ denote the set of all supervised positions included in the training objective. For each
$q\in\mathcal Q$, let $\tau_q\in\mathcal V$ denote the correct target token and
let $\mathbf h_q^{(t)}\in\R^{d_{\mathrm{model}}}$ denote the final-layer
hidden state used to predict $\tau_q$. The output logits, probabilities, and
summed supervised loss are then given by
\begin{equation}
    z_{q,u}^{(t)} = \bigl(\mathbf{h}_q^{(t)}\bigr)^{\trans}\mathbf{o}_u^{(t)}, \quad p_{q,u}^{(t)} = \frac{\exp\bigl(z_{q,u}^{(t)}\bigr)}{\sum_{w\in\mathcal{V}}\exp\bigl(z_{q,w}^{(t)}\bigr)}, \quad \mathcal L^{(t)} = -\sum_{q\in\mathcal Q}\log p_{q,\tau_q}^{(t)}.
    \label{eq:vector-full-loss}
\end{equation}
Additionally, for a level-$\ell$ code $v$, define its supervision count as
\(
    n_v=\sum_{q\in\mathcal Q}\mathbb{1}[{\tau_q=v}]
\)
and let $N_\ell=\sum_{u=0}^{K-1}n_u>0$ denote the total number of
level-$\ell$ targets. The average and mean-centered counts for $v$ are
\begin{equation}
    \bar n=\frac{N_\ell}{K}, \qquad
    \widetilde n_v=n_v-\bar n.
    \label{eq:vector-centered-count}
\end{equation}
As each occurrence of item $i$ contributes its level-$\ell$ code
$k_\ell(i)$ as a target,
$n_v=\sum_i f_i^{\mathcal D}\mathbb{1}[{k_\ell(i)=v}]$,
where $f_i^{\mathcal D}$ denotes the number of times item $i$ appears as a
supervised target in the training corpus $\mathcal D$.
Thus, $n_v$ is the aggregate popularity of the items assigned
to code $v$.

\vspace{0.1cm}
\noindent\textbf{Gradient with Respect to $\mathbf{o}_v^{(t)}$.}
We begin by deriving the gradient of the loss with respect to the output
embedding $\mathbf{o}_v^{(t)}$. For a supervised position
$q\in\mathcal Q$, the position-level loss can be written as
\begin{equation}
    \mathcal L_q^{(t)}
    =-\log p_{q,\tau_q}^{(t)}
    =-\bigl(\mathbf h_q^{(t)}\bigr)^{\trans}
       \mathbf o_{\tau_q}^{(t)}
     +\log\sum_{w\in\mathcal V}
       \exp\left(
       \bigl(\mathbf h_q^{(t)}\bigr)^{\trans}
       \mathbf o_w^{(t)}
       \right).
    \label{eq:vector-position-loss}
\end{equation}
Since
$\mathbf{o}_v^{(t)}$ only impacts the loss through the output logits, the hidden state $\mathbf h_q^{(t)}$ can be treated as fixed when
differentiating with respect to $\mathbf{o}_v^{(t)}$. As such, the gradient
contribution from position $q$ is
\[
    \frac{\partial\mathcal L_q^{(t)}}
         {\partial\mathbf o_v^{(t)}}
    =
    \left(
        p_{q,v}^{(t)}-\mathbb{1}[{\tau_q=v}]
    \right)\mathbf h_q^{(t)}.
\]
As $\mathcal L^{(t)}=\sum_{q\in\mathcal Q}\mathcal L_q^{(t)}$,
linearity of differentiation gives
\begin{equation}
    \mathbf g_v^{(t)}
    =
    \frac{\partial\mathcal L^{(t)}}
         {\partial\mathbf o_v^{(t)}}
    =
    \sum_{q\in\mathcal Q}
    \left(
        p_{q,v}^{(t)}-\mathbb{1}[{\tau_q=v}]
    \right)\mathbf h_q^{(t)}.
    \label{eq:vector-row-gradient}
\end{equation}

\vspace{0.1cm}
\noindent\textbf{Gradient Descent Step.}
For each level-$\ell$ code $v$, define
\begin{align}
    \mathbf s_v^{(t)}
    &=
    \sum_{\substack{q\in\mathcal Q\\ \tau_q=v}}
    \mathbf h_q^{(t)}, \nonumber\\
    \mathbf a_v^{(t)}
    &=
    \sum_{q\in\mathcal Q}
    p_{q,v}^{(t)}\mathbf h_q^{(t)}.
    \label{eq:vector-gradient-components}
\end{align}
Here, $\mathbf s_v^{(t)}$ aggregates the hidden states at positions where
$v$ is the correct target, whereas $\mathbf a_v^{(t)}$ aggregates the
hidden states across all supervised positions according to the probability
assigned to $v$. Thus,
\[
    \mathbf g_v^{(t)}
    =
    \mathbf a_v^{(t)}-\mathbf s_v^{(t)}.
\]

Given we are interested in the relative geometry among the $K$
level-$\ell$ output rows, which determines their pairwise logit differences
for a fixed hidden state, define
\begin{align}
    \bar{\mathbf o}^{(t)}
    &=\frac{1}{K}\sum_{u=0}^{K-1}\mathbf o_u^{(t)},
    &
    \mathbf x_v^{(t)}
    &=\mathbf o_v^{(t)}-\bar{\mathbf o}^{(t)}, \nonumber\\
    \bar{\mathbf a}^{(t)}
    &=\frac{1}{K}\sum_{u=0}^{K-1}\mathbf a_u^{(t)},
    &
    \bar{\mathbf s}^{(t)}
    &=\frac{1}{K}\sum_{u=0}^{K-1}\mathbf s_u^{(t)}.
    \label{eq:vector-centered-quantities}
\end{align}

For a learning rate $\eta>0$, gradient descent updates each output row as
\[
    \mathbf o_v^{(t+1)}
    =
    \mathbf o_v^{(t)}-\eta\mathbf g_v^{(t)}.
\]
Averaging this update over the $K$ level-$\ell$ rows gives
\[
    \bar{\mathbf o}^{(t+1)}
    =
    \bar{\mathbf o}^{(t)}
    -\frac{\eta}{K}
    \sum_{u=0}^{K-1}\mathbf g_u^{(t)}.
\]
Subtracting the mean update from the update for row $v$ gives
\begin{align}
    \mathbf x_v^{(t+1)}
    &=
    \mathbf x_v^{(t)}
    -\eta
    \left[
        \mathbf g_v^{(t)}
        -\frac{1}{K}\sum_{u=0}^{K-1}\mathbf g_u^{(t)}
    \right]
    \nonumber\\
    &=
    \mathbf x_v^{(t)}
    +\eta
    \left(
        \mathbf s_v^{(t)}-\bar{\mathbf s}^{(t)}
    \right)
    -\eta
    \left(
        \mathbf a_v^{(t)}-\bar{\mathbf a}^{(t)}
    \right).
    \label{eq:vector-centered-gradient-step}
\end{align}
This identity shows that only code-specific differences in the target and
probability terms affect the relative geometry, i.e., any update shared by all
code rows cancels under centering.

To isolate the target-frequency contribution, define
\begin{equation}
    \bar{\mathbf h}^{(t)}
    =
    \frac{1}{N_\ell}
    \sum_{u=0}^{K-1}\mathbf s_u^{(t)},
    \qquad
    \mathbf d_v^{(t)}
    =
    \mathbf s_v^{(t)}
    -n_v\bar{\mathbf h}^{(t)}.
    \label{eq:vector-context-decomposition}
\end{equation}
Because
\[
    \bar{\mathbf s}^{(t)}
    =
    \frac{N_\ell}{K}\bar{\mathbf h}^{(t)}
    =
    \bar n\bar{\mathbf h}^{(t)},
\]
we have
\begin{equation}
    \mathbf s_v^{(t)}-\bar{\mathbf s}^{(t)}
    =
    \widetilde n_v\bar{\mathbf h}^{(t)}
    +\mathbf d_v^{(t)}.
    \label{eq:vector-centered-target}
\end{equation}
The first term isolates an explicit contribution of $v$'s relative target
frequency,
while $\mathbf d_v^{(t)}$ captures how its target contexts differ from the
codebook average.
Similarly, define the centered probability contribution as
\begin{align}
    \mathbf R_v^{(t)}
    &=
    \mathbf a_v^{(t)}-\bar{\mathbf a}^{(t)}
    \nonumber\\
    &=
    \sum_{q\in\mathcal Q}
    \left(
        p_{q,v}^{(t)}-\bar p_q^{(t)}
    \right)\mathbf h_q^{(t)},
    \qquad
    \bar p_q^{(t)}
    =
    \frac{1}{K}\sum_{u=0}^{K-1}p_{q,u}^{(t)}.
    \label{eq:vector-centered-probability}
\end{align}
Substituting these decompositions into
\Cref{eq:vector-centered-gradient-step} yields
\begin{equation}
    \mathbf x_v^{(t+1)}
    =
    \mathbf x_v^{(t)}
    +\eta\widetilde n_v\bar{\mathbf h}^{(t)}
    +\eta\boldsymbol\Delta_v^{(t)},
    \qquad
    \boldsymbol\Delta_v^{(t)}
    =
    \mathbf d_v^{(t)}-\mathbf R_v^{(t)}.
    \label{eq:vector-exact-centered-step}
\end{equation}
Here, the first update term depends explicitly on the
relative target frequency of $v$, i.e., its relative popularity, while the residual retains variation in
both its target contexts and its predicted probabilities.

\vspace{0.1cm}
\noindent \textbf{Accumulation Across Steps}
Under the full-batch setting of \Cref{thm:accumulated-bias-main}, the training
positions and supervision counts remain fixed across steps. In particular,
$\widetilde n_v$ does not depend on $t$, although the hidden states,
probabilities, and residual generally evolve as the model is updated.
Rearranging \Cref{eq:vector-exact-centered-step} and summing from $t=0$ to
$S-1$ gives
\begin{align}
    \sum_{t=0}^{S-1}
    \left(
        \mathbf x_v^{(t+1)}-\mathbf x_v^{(t)}
    \right)
    &=
    \eta\widetilde n_v
    \sum_{t=0}^{S-1}\bar{\mathbf h}^{(t)}
    +
    \eta\sum_{t=0}^{S-1}\boldsymbol\Delta_v^{(t)}.
    \label{eq:vector-summed-step}
\end{align}
The left-hand side simplifies to
$\mathbf x_v^{(S)}-\mathbf x_v^{(0)}$. Therefore,
\begin{equation}
    \mathbf x_v^{(S)}
    =
    \mathbf x_v^{(0)}
    +
    \eta\widetilde n_v
    \sum_{t=0}^{S-1}\bar{\mathbf h}^{(t)}
    +
    \eta\sum_{t=0}^{S-1}\boldsymbol\Delta_v^{(t)}.
    \label{eq:vector-accumulated-step}
\end{equation}
In aggregate, the second term identifies an
explicit frequency-dependent component where every code is shifted along the
shared direction $\sum_{t=0}^{S-1}\bar{\mathbf h}^{(t)}$ by an amount
proportional to its relative popularity $\widetilde n_v$. 
\hfill $\square$

\vspace{0.1cm}
\noindent\textit{Remark on the Initial Residual.}
Suppose the untied output rows are initialized independently using the rule in
\Cref{eq:randinit}. As these rows are identically distributed, no level-\(\ell\) code is systematically assigned a probability above or below the codebook average when conditioned on the hidden states. Formally,
\[
    \mathbb E_{\mathrm{init}}\!\left[
        p_{q,v}^{(0)}-\bar p_q^{(0)}
        \;\middle|\;
        \{\mathbf h_q^{(0)}\}_{q\in\mathcal Q}
    \right]
    =0.
\]
Accordingly, $\mathbf R_v^{(0)}$, which aggregates these centered probability
differences, is conditionally zero-mean over random initializations.

Random initialization further motivates, but does not guarantee, an
early-stage context-homogeneity approximation. As the model has not yet
learned distinct semantic representations for the SID tokens, the mean hidden
states at positions targeting different codes may be approximately
code-independent. Under this approximation,
\[
    \mathbf s_v^{(0)}
    \approx
    n_v\bar{\mathbf h}^{(0)},
    \qquad\text{and hence}\qquad
    \mathbf d_v^{(0)}
    \approx
    \mathbf 0.
\]
Although context variations across different codes may weaken this approximation in practice, it reflects a natural symmetry at initialization. Notably, our proposal of centroid initialization is designed to break this symmetry by supplying code-specific semantic structure at initialization. Note that these symmetry arguments apply exclusively to the initial residual and do not constrain $\boldsymbol\Delta_v^{(t)}$ as training progresses.
\section{Additional Metric Results}
\label{app:recall10}

\begin{figure}[t]
\centering
\includegraphics[width=0.43\textwidth]{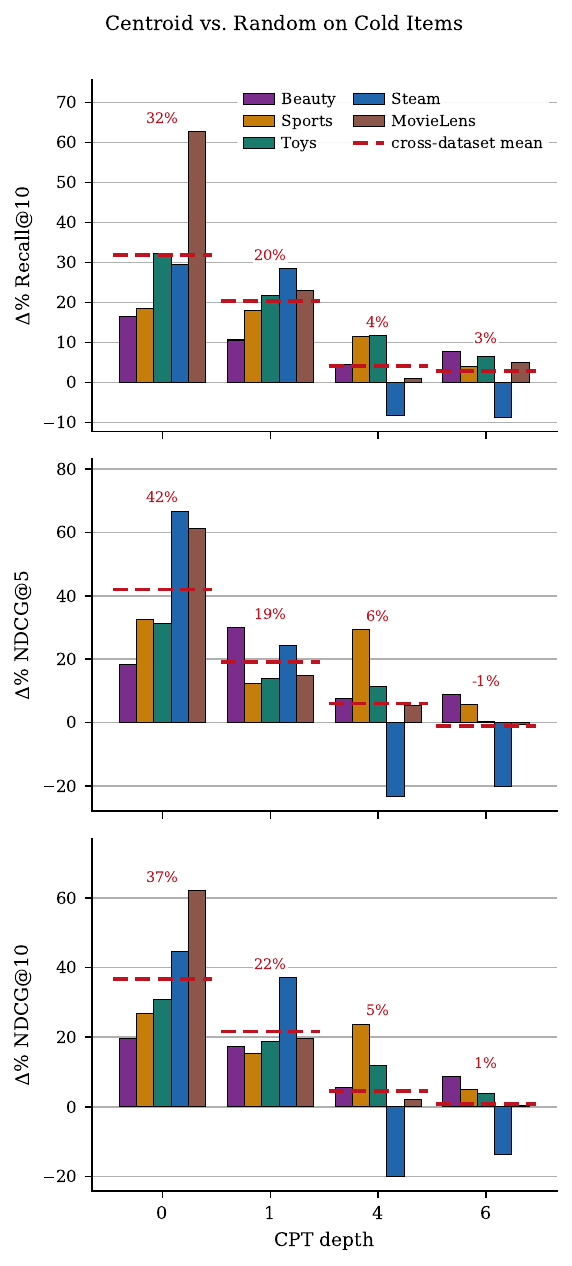}
\caption{Relative metric gaps (centroid $-$ random) on cold targets (train-target freq.\ $\leq\!1$) versus the number of CPT epochs. With fewer CPT epochs, centroid initialization tends to produce larger cold-item metric gains.}
\label{fig:tierdepth_cold_app}
\end{figure}

To complement the main paper results, in this section we provide additional results for Recall@10, NDCG@5, and NDCG@10 to the Recall@5 results in \Cref{fig:cptdepth} and \Cref{fig:tierdepth_cold}. Additionally, we report NDCG@5 and NDCG@10 for \Cref{tab:av1}. 

\vspace{0.1cm}
\noindent \textbf{Pure-SFT NDCG Metrics.} In \Cref{tab:av1-ndcg}, we report NDCG across the five datasets from the pure-SFT experiments. Consistent with the recall improvements observed in the main text, centroid initialization consistently outperforms the random baseline across all datasets for both NDCG@5 and NDCG@10. Notably, because NDCG is a rank-sensitive metric, these gains demonstrate that our semantic prior actively pushes the ground-truth SID higher up the ranked list, rather than merely surfacing it within the top-$k$. For example, on the Toys and Sports datasets, centroid initialization yields relative NDCG@5 improvements of 11.1\% and 9.5\%, respectively, showing that preserving the underlying semantic geometry translates directly into a higher-quality ranking distribution.

\begin{figure*}[t!]
\centering
\includegraphics[width=0.98\textwidth]{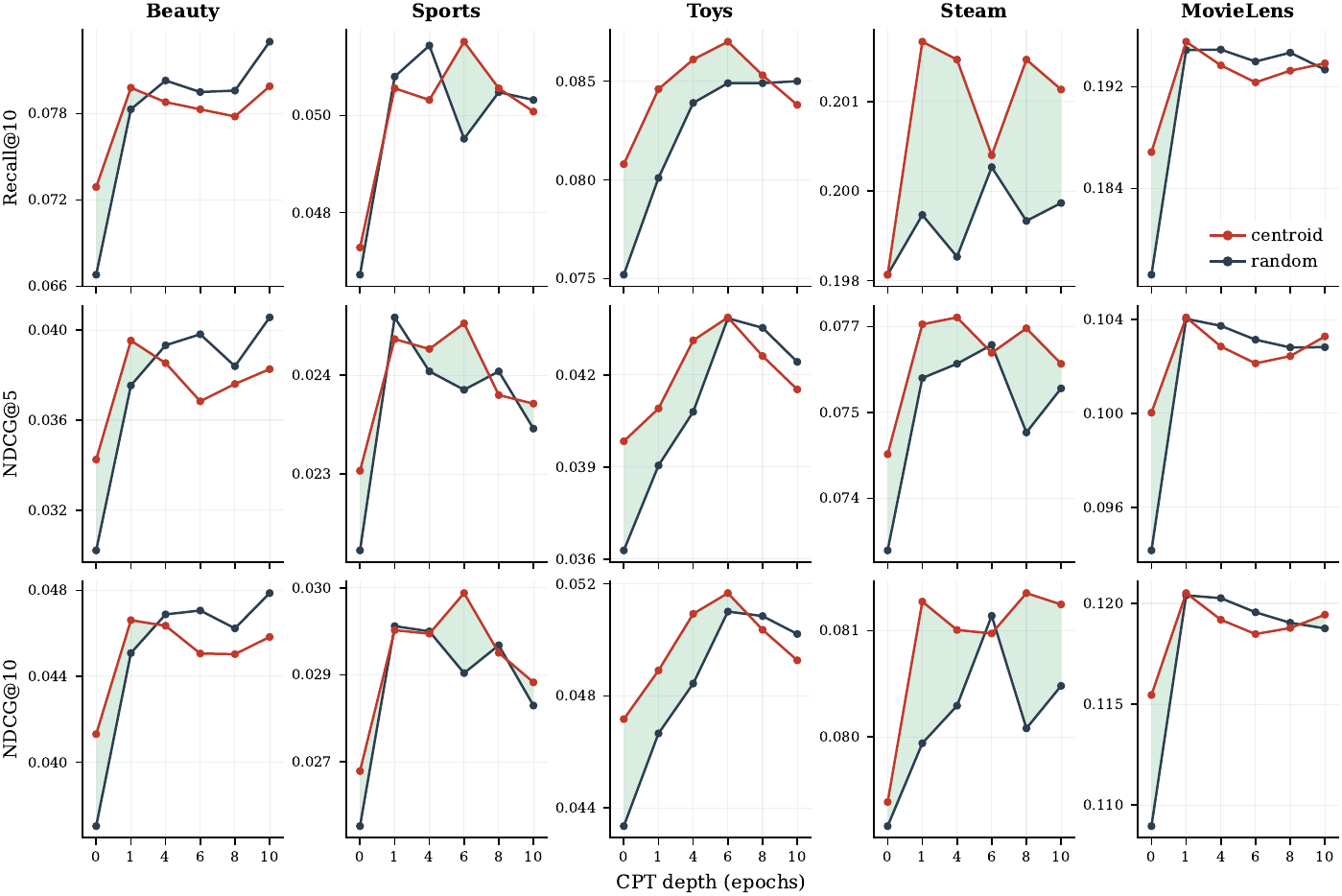}
\caption{
Recall@10, NDCG@5, and NDCG@10 across numbers of CPT epochs. Centroid initialization often reaches the random-init performance with fewer CPT epochs across all metrics, supporting the findings of the main text.}
\label{fig:cptdepth_app}
\end{figure*}

\vspace{0.1cm}
\noindent \textbf{Recall@10 and NDCG Across CPT Epochs.} In \Cref{fig:cptdepth_app}, we expand our CPT-epoch analysis to include Recall@10, NDCG@5, and NDCG@10. Consistent with the Recall@5 trends observed in the main text, centroid initialization generally accelerates convergence to peak performance across the wide range of metrics with fewer CPT epochs. For example, on the Toys, Steam, and MovieLens datasets, the centroid-initialized models achieve strong NDCG and Recall@10 metrics with just 1 to 4 CPT epochs, whereas the random baseline tends to require additional epochs to reach similar performance. Sports is the one exception where centroid initialization yields a higher pure-SFT baseline at zero epochs but requires additional CPT epochs to perfectly match the random baseline's peak. That said, the absolute performance differences in Sports at early CPT epochs are relatively small, often differing only in the fourth decimal place, suggesting that the practical impact of this delayed convergence is minimal. 
\vspace{0.1cm}

\noindent \textbf{Cold Item Recall@10 and NDCG.} In \Cref{fig:tierdepth_cold_app}, we extend our cold-item analysis to include Recall@10, NDCG@5, and NDCG@10. Consistent with the Recall@5 trends observed in the main text, centroid initialization yields large relative gains in the pure-SFT regime, averaging 32\% for Recall@10, 42\% for NDCG@5, and 37\% for NDCG@10 across the five datasets. The notably high improvement in NDCG@5 (42\%) suggests that the semantic prior not only successfully retrieves cold items, but also actively pushes them to the top of the ranked list. Furthermore, we observe that as the number of CPT epochs increases, the random baseline slowly aligns with the text space, causing the relative advantage to diminish. Even so, after one CPT epoch, the cross-dataset mean advantages remain strong at 20\% for Recall@10, 19\% for NDCG@5, and 22\% for NDCG@10, before eventually converging near parity (3\%, -1\%, and 1\%, respectively) by epoch 6. Ultimately, these metrics corroborate our earlier findings that centroid initialization provides strong tail-item generalization, largely bypassing the need for costly CPT epochs to establish cold-item semantics.

\begin{table}[]
\centering
\caption{\textbf{NDCG counterpart of \Cref{tab:av1} (pure-SFT).} NDCG@\{5,10\} with gain
$1/\log_2(\mathrm{rank}+1)$ for the ground-truth item. \textcolor[HTML]{5315f5}{\textbf{Bold}}
is better.}
\label{tab:av1-ndcg}
\begin{tabular}{llcc}
\toprule
Domain & Initialization & NDCG@5 & NDCG@10 \\
\midrule
\multirow{2}{*}{Beauty}
& Random   & 0.030 & 0.037 \\
& Centroid & \textbf{\textcolor[HTML]{5315f5}{0.034}} & \textbf{\textcolor[HTML]{5315f5}{0.041}} \\
\midrule
\multirow{2}{*}{Sports}
& Random   & 0.021 & 0.026 \\
& Centroid & \textbf{\textcolor[HTML]{5315f5}{0.023}} & \textbf{\textcolor[HTML]{5315f5}{0.027}} \\
\midrule
\multirow{2}{*}{Toys}
& Random   & 0.036 & 0.043 \\
& Centroid & \textbf{\textcolor[HTML]{5315f5}{0.040}} & \textbf{\textcolor[HTML]{5315f5}{0.047}} \\
\midrule
\multirow{2}{*}{Steam}
& Random   & 0.073 & 0.078 \\
& Centroid & \textbf{\textcolor[HTML]{5315f5}{0.074}} & \textbf{\textcolor[HTML]{5315f5}{0.079}} \\
\midrule
\multirow{2}{*}{MovieLens}
& Random   & 0.094 & 0.109 \\
& Centroid & \textbf{\textcolor[HTML]{5315f5}{0.100}} & \textbf{\textcolor[HTML]{5315f5}{0.115}} \\
\bottomrule
\end{tabular}
\end{table}

\section{Additional Ablations}

To isolate the geometric properties of the centroid table from the specific code to centroid assignment it encodes, we evaluate a permuted initialization where the 256 level-$\ell$ centroids are shuffled among the 256 level-$\ell$ codes using a fixed random permutation. As this is a permutation, all aggregate marginal properties of the initialization (i.e., per-level row norms, per-level means, and the mean-shift frame of \Cref{eq:centinit}) are preserved, while the semantic correspondence between a code and its content is removed. Given the change in behavior between a CPT depth of 1 and 10, seen in the main text, we study both of these settings here as well (denoted $d1$ and $d10$). 

As shown in \Cref{tab:abl-shuffled}, this loss of semantic alignment consistently degrades performance. Specifically, eight of the settings suffer a drop in recall (reaching $-5.7\%$ on Sports CPT+SFT $d1$), alongside some increases in convergence times (e.g., $+3$ epochs on Beauty pure-SFT). Although these aggregate metrics appear modest, we find that the embedding diagnostics reveal deeper structural problems. First, we find that the trained SID embeddings barely deviate from their shuffled initializations, e.g., their similarity is $\geq 0.94$ against the shuffled table, but $\approx 0$ against the true centroids. On the surface, this appears beneficial, as the embeddings successfully retain a well-conditioned spectrum and avoid the popularity collapse characteristic of random initialization (pop-$R^2@3$ = $0.01$). However, these embeddings also carry little category signal (cat-$R^2@3$ = $0.01$).
Consequently, on cold targets, we find that shuffling tends to remove the benefits of the centroid structure, causing performance on Beauty to fall below random initialization ($0.012$ for shuffled vs. $0.023$ for the canonical centroid and $0.016$ for random). Likewise, this degradation scales with item rarity, yielding relative performance factors of $\times 0.97$ for head items, $\times 0.85$ for tail items, and $\times 0.53$ for cold items. While a similar, but less severe, degradation occurs on Toys ($\times 0.89$ on cold items), the direction is consistent. Together, these results demonstrate that centroid initialization confers two distinct benefits: a well-conditioned, high-rank row geometry that does not encode popularity, and a code to content correspondence that benefits cold-item retrieval.

\begin{table}[t]
\centering
\caption{\textbf{Shuffled-centroid ablation.} We measure the impact on recall ($\Delta$\%) and convergence time (difference in number of training epochs before early stopping, measured as $\Delta$ep) when shuffling code-to-centroid assignments. Shuffling consistently degrades both metrics (\textcolor[HTML]{5315f5}{\textbf{bold}} indicates degraded recall or increased training time), demonstrating that semantic alignment is beneficial.}
\label{tab:abl-shuffled}
\begin{tabular}{llcccc}
\toprule
 & & \multicolumn{2}{c}{Orig. Centroid} & \multicolumn{2}{c}{Shuffled Centroid} \\
\cmidrule(lr){3-4}\cmidrule(lr){5-6}
Domain & Stage & Recall@5 & Ep. & $\Delta$\% & $\Delta$Ep. \\
\midrule
\multirow{3}{*}{Beauty}
 & pure-SFT      & 0.051 & 2 & \textbf{\textcolor[HTML]{5315f5}{-5.3\%}} & \textbf{\textcolor[HTML]{5315f5}{+3}} \\
 & CPT+SFT $d1$  & 0.057 & 4 & \textbf{\textcolor[HTML]{5315f5}{-0.5\%}} & +0 \\
 & CPT+SFT $d10$ & 0.057 & 4 & \textbf{\textcolor[HTML]{5315f5}{-2.8\%}} & -1 \\
\addlinespace
\multirow{3}{*}{Sports}
 & pure-SFT      & 0.033 & 5 & \textbf{\textcolor[HTML]{5315f5}{-3.6\%}} & +0 \\
 & CPT+SFT $d1$  & 0.036 & 3 & \textbf{\textcolor[HTML]{5315f5}{-5.7\%}} & +0 \\
 & CPT+SFT $d10$ & 0.035 & 3 & \textbf{\textcolor[HTML]{5315f5}{-0.5\%}} & +0 \\
\addlinespace
\multirow{3}{*}{Toys}
 & pure-SFT      & 0.058 & 5 & \textbf{\textcolor[HTML]{5315f5}{-1.8\%}} & +0 \\
 & CPT+SFT $d1$  & 0.059 & 4 & \textbf{\textcolor[HTML]{5315f5}{-3.3\%}} & \textbf{\textcolor[HTML]{5315f5}{+1}} \\
 & CPT+SFT $d10$ & 0.060 & 4 & +1.2\% & -1 \\
\bottomrule
\end{tabular}
\end{table}
\end{document}